\documentclass[twocolumn]{aastex6}

\usepackage{amsmath}
\usepackage{amssymb}
\usepackage{float}
\usepackage{hyperref}
\usepackage{nicefrac}
\usepackage{xcolor}
\usepackage{cancel}
\usepackage[capitalise]{cleveref}
\usepackage{booktabs}

\newcommand{\nCEFT}{n_\text{EFT}}
\newcommand{\pCEFT}{p_\text{EFT}}
\newcommand{\nsat}{n_0}
\newcommand{\n}{n}

\newcommand{\nPT}{n_\text{PT}}
\newcommand{\Dn}{\Delta n}
\newcommand{\ePT}{\epsilon_\text{PT}}
\newcommand{\De}{\Delta \epsilon}

\newcommand{\eos}{EOS}
\newcommand{\eoss}{EOSs}

\makeatletter
\g@addto@macro\bfseries{\boldmath}
\makeatother

\usepackage{graphicx}

\begin{document}

\title{Constraints on strong phase transitions in neutron stars}

\slugcomment{HIP-2022-34/TH}

\author{T.\ Gorda,$^{1,2}$ K.\ Hebeler,$^{1,2,3}$ A.\ Kurkela,$^4$ A.\ Schwenk$\,^{1,2,3}$ and A.\ Vuorinen$^5$}
\affil{
$^1$Technische Universit\"at Darmstadt, Department of Physics, 64289 Darmstadt, Germany \\
$^2$ExtreMe Matter Institute EMMI, GSI Helmholtzzentrum f\"ur Schwerionenforschung GmbH, 64291 Darmstadt, Germany \\
$^3$Max-Planck-Institut f\"ur Kernphysik, Saupfercheckweg 1, 69117 Heidelberg, Germany \\
$^4$Faculty of Science and Technology, University of Stavanger, 4036 Stavanger, Norway\\
$^5$Department of Physics and Helsinki Institute of Physics, P.O.~Box 64, FI-00014 University of Helsinki, Finland
}

\begin{abstract}
We study current bounds on strong first-order phase transitions (PTs) along the equation of state (\eos) of dense strongly interacting matter in neutron stars, under the simplifying assumption that on either side of the PT the \eos\ can be approximated by a simple polytropic form. We construct a large ensemble of possible \eoss\ of this form, anchor them to chiral effective field theory calculations at nuclear density and perturbative QCD at high densities, and subject them to astrophysical constraints from high-mass pulsars and gravitational-wave observations. Within this setup, we find that a PT permits neutron-star solutions with larger radii, but only if the transition begins below twice nuclear saturation density. We also identify a large parameter space of allowed PTs currently unexplored by numerical-relativity studies. Additionally, we locate a small region of parameter space allowing twin-star solutions, though we find them to only marginally pass the current astrophysical constraints. Finally, we find that sizeable cores of high-density matter beyond the PT may be located in the centers of some stable neutron stars, primarily those with larger masses.
\end{abstract}

\maketitle

\section{Introduction}
\label{sec:Intro}

The detailed phase structure of Quantum Chromodynamics (QCD), the theory governing the strong nuclear interaction, is still largely unknown. A combination of lattice-field-theory calculations and two decades of ultra-relativistic heavy-ion-collision experiments carried out at RHIC and at the LHC has established the crossover nature of the deconfinement transition at vanishing and small baryon number chemical potentials \citep{Aoki:2006we,Cheng:2006qk}. However, the phase structure of baryon-rich matter present in the cores of neutron stars (NSs), and probed in lower-energy heavy-ion collisions, remains uncharted. 

Recent years have witnessed a rapid evolution in NS observations, which has for the first time permitted model-independent constraints of the equation of state (\eos) of the dense matter in NSs. This inference of the \eos\ has been based on the generation of large ensembles of model-agnostic \eoss\ that are then conditioned to be consistent with both NS observations and \textit{ab initio} calculations \citep{Hebeler:2013nza,Kurkela:2014vha,Most:2018hfd,Annala:2017llu, Tews:2018iwm,Landry:2018prl,Capano:2019eae,Miller:2019nzo,Essick:2019ldf,Raaijmakers:2019dks,Annala:2019puf,Dietrich:2020efo,Landry:2020vaw,Al-Mamun:2020vzu,Miller:2021qha,Essick:2021kjb,Raaijmakers:2021uju,
Annala:2021gom,Huth:2021bsp,Altiparmak:2022bke,Lim:2022fap,Gorda:2022jvk}. The observations that currently constrain the \eos\ the most include the existence of massive NSs with masses around and exceeding the two-solar-mass limit \citep{Demorest:2010bx,Antoniadis:2013pzd,NANOGrav:2019jur,Fonseca:2021wxt}, constraints on the tidal deformability obtained from the binary-NS merger GW170817 \citep{LIGOScientific:2017vwq,LIGOScientific:2018cki,LIGOScientific:2018hze}, as well as the simultaneous mass-radius measurements performed with the NICER telescope \citep{Miller:2019cac,Riley:2019yda,Miller:2021qha,Riley:2021pdl}. On the theoretical side, the \eos\ is constrained by \textit{ab initio} calculations based on chiral effective field theory (EFT) at nuclear densities \citep{Hebeler:2013nza,Tews:2012fj,Lynn:2015jua,Drischler:2017wtt,Drischler:2020hwi,Keller:2022crb} and perturbative QCD (pQCD) at very high densities \citep{Kurkela:2009gj, Gorda:2021kme}. 

While the \eos\ is an interesting and fundamental quantity in itself, interest in the \eos\ also arises from the fact that its features can reflect the active degrees of freedom in the system and hence the physical phase of matter. In the past years, there have been several works that have suggested features of the \eos\ may be related to the onset of hyperonic matter [e.g., \citep{Gal:2016boi}], the onset of quark matter [e.g., \citep{Annala:2019puf}], the presence of a quarkyonic phase \citep{McLerran:2018hbz} or of diquark pairing [e.g., \citep{Leonhardt:2019fua}]. The most dramatic connection between the \eos\ and the phase structure takes place if the \eos\ contains a strong first-order phase transition (PT) between a low-density hadronic phase and a new high-density phase.

Indeed, there have been multiple studies of \eoss\ featuring strong PTs, displaying, e.g., qualitatively different gravitational-wave signals arising from binary-NS mergers compared to a scenario with vanishing latent heat \citep{Most:2018eaw, Bauswein:2018bma, Fujimoto:2022xhv}. These studies have, however, been restricted to individual models, so it is of great interest to systematically determine what bounds on first-order PTs follow from currently existing theoretical and observational constraints. Such results would be of great value for merger simulations, as they would map out the space of possible PTs and suggest what ranges of \eoss\ one should cover. 

While very versatile, the model-agnostic \eos-inference studies performed so far have focused less on exploring the broad space of strong PTs. In this work, our objective is to perform a simple inference study of the NS-matter \eos\ including first-order PTs and setting no \textit{a priori} limitations for the strength and onset density of the transition. We do so by generating a large ensemble of \eoss\ that continue from a chiral EFT \eos\ band using a two-segment polytrope with a PT inserted between them. This \eos\ is then constructed up to densities well beyond those reached in stable NSs, in practice ten times nuclear saturation densities $n = 10 \nsat$ (with $\nsat = 0.16$~fm$^{-3}$). At this density two methods, detailed in Sec.~\ref{sec:Setup}, are used to restrict the range of allowed values for the energy density $\epsilon$ and pressure $p$ to enforce a high-density constraint from pQCD. This simple setup, though less general than previous model-agnostic approaches without a first-order PT, allows us to clearly delineate the matter below and above the transition and remains flexible enough to allow a general exploration of PTs along the NS-matter \eos\ without resorting to a specific microscopic model at intermediate densities.  

This article is structured as follows. Following the details of our \eos\ construction in Sec.~\ref{sec:Setup}, we divide the discussion into smaller sections, each focusing on specific findings. First, we investigate the effects of a PT on the set of possible masses and radii of stable NSs in Sec.~\ref{sec:MR_discussion}, finding larger-radius stars than were found in previous model-agnostic studies. In Sec.~\ref{sec:general_matching}, we then study the range of allowed PT parameters, namely the onset densities and latent heats consistent with current NS observations and \textit{ab initio} calculations. Here we also compare these allowed ranges with specific models used in the merger literature and investigate the parameter space for twin-star (or third-family) solutions within our approach. The latter are discussed in detail in Sec.~\ref{sec:twins}. Finally, we investigate the possibility of cores of high-density matter within stable massive NSs in Sec.~\ref{sec:cores}, detailing the ranges of NS masses and radii consistent with them. Finally, we conclude with a summary in Sec.~\ref{sec:summary}.

\section{Setup}
\label{sec:Setup}

We build our \eos\ ensemble using a continuous piecewise construction up to the baryon density of $10 \nsat$. For the density region $n \leq 0.57 \nsat$, we use the BPS crust \eos\ from \citet{Baym:1971pw}, followed by an \eos\ within the chiral EFT band spanned by the ``stiff'' or ``soft'' \eoss\ from \citet{Hebeler:2013nza} up to a density $\nCEFT \equiv 1.1 \nsat$. For simplicity, we generate intermediate \eoss\ between the stiff and soft \eoss\ with a linear interpolation of $p$, $\epsilon$, and baryon chemical potentials at fixed $n$, which well approximates the results obtained from many-body calculations in that work.
 Beyond this density, we extend the \eos\ with a two-polytrope construction with a first-order PT in between. The parameters of the PT, $\nPT$ and $\Dn$, specify the onset number density and jump in number density at the PT, respectively. Explicitly, this implies: 
\begin{enumerate}
    \item For $\nCEFT < n \leq \nPT$, the \eos\ is governed by a single polytropic form, namely, $p(n) = \pCEFT(\nCEFT) (n/\nCEFT)^{\Gamma_1}$,
    \item For $\nPT < n \leq \nPT+\Dn$, there is a first-order PT where the baryon density jumps by $\Delta n$ but the pressure remains constant,
    \item For $\nPT + \Dn < n \leq 10n_0$, we use another polytropic \eos, matched to the end of the PT: $p(n) = p(\nPT) [n/(\nPT + \Dn)]^{\Gamma_2}$.
\end{enumerate}
At the density of $10\nsat$ we use one of two possible high-density constraints. First, we use the \eos\ ensemble generated in \citet{Annala:2021gom} to discard all \eoss\ with inconsistent {$\epsilon$, $p$} values at this density. This choice allows us to make a comparison to this work, whose ensemble did not feature explicit PTs, and therefore to directly assess the effect of first-order PTs on such \eos\ ensembles and the macroscopic properties of the corresponding NSs.
Second, we match to the less restrictive $10 \nsat$ region of \citet{Komoltsev:2021jzg}, which is obtained by demanding that the extrapolated \eoss\ can be connected to the perturbative QCD \eos\ at densities around $40\nsat$ (where this calculation is under quantitative control). We use this second matching when comparing the resulting allowed range of PTs with the models that have been used in recent binary-NS simulations. 

In all, our construction requires five parameters to fully specify a given \eos:
\begin{enumerate}
    \item a continuous parameter $s \in [0,1]$ that linearly interpolates between the soft and stiff chiral EFT \eoss\ of \citet{Hebeler:2013nza} at fixed $n$, 
    \item $\nPT \in [1.1, 10]\nsat$,
    \item $\Dn/\nPT \in [0,9.09]$,
    \item $\epsilon$ within the \citet{Annala:2021gom} or \citet{Komoltsev:2021jzg} regions at $10\nsat$,
    \item $p$ within the same region at $10\nsat$.
\end{enumerate}
These five parameters are sampled from uniform distributions and together uniquely fix $\Gamma_1$ and $\Gamma_2$. 
Explicitly, $s$, $\Gamma_1$, $\nPT$, $\Dn/\nPT$, and $\Gamma_2$ uniquely fix $\epsilon, p$ at $10 \nsat$, and this process can be inverted.
Note that the upper bound on $\Dn/\nPT$ is a consequence of the construction: any \eos\ is constrained to have $\Dn \leq 10\nsat - \nCEFT$. We additionally reject any \eos\ that becomes superluminal (with an acausal speed of sound $c_s^2 > 1$) in the interpolated region. The five parameters are sampled uniformly, and for the present study, we sample in total about 1,500,000 possible causal \eoss. 

Once we have generated our ensemble of \eoss, two astrophysical constraints are folded in as hard cuts: 
\begin{enumerate}
    \item Any viable \eos\ must be able to support a NS of $2 M_\odot$ mass, to be compatible with the observations of high-mass NSs.\footnote{Note that this constraint is rather conservative, as several NSs have been recently identified with masses likely exceeding this limit \citep{Linares:2018ppq,NANOGrav:2019jur,Fonseca:2021wxt,Romani:2022jhd}. Similarly, we note that our way of implementing the tidal-deformability constraint is on purpose rather conservative and could in principle be slightly strengthened.}
    \item Any viable \eos\ must be consistent with the LIGO/Virgo 90\% credible interval for the tidal deformability $\Lambda$ inferred from the GW170817 event.
    Here, we use the fact that the chirp mass is known to very good accuracy $\mathcal{M}_\mathrm{chirp} = 1.186 M_\odot$, while the mass ratio and tidal deformability are constrained by $q = M_2/M_1 > 0.73$ and $\tilde{\Lambda} < 720$
    \citep{LIGOScientific:2018hze}, with the latter defined by
     \begin{equation}
        \hspace*{10mm} \tilde{\Lambda} = \frac{16}{13} \left[ \frac{(M_1 + 12 M_2) M_1^4 \Lambda(M_1)}{(M_1+M_2)^5} + (M_1 \leftrightarrow M_2) \right]. \nonumber
    \end{equation}
    In this relation, $M_1$ and $M_2$ are the gravitational masses of the two NSs in the event  (with $M_2<M_1$), 
    and $\Lambda(M_1)$ and $\Lambda(M_2)$ are their respective tidal deformabilities (assuming a common \eos) as defined in~\citet{Hinderer:2007mb}.
\end{enumerate}
After implementing these two constraints, approximately 280,000 \eoss\ of our sample remain as viable candidate \eoss\ containing a first-order PT. This is the ensemble that we proceed to systematically investigate.

\begin{figure}[t]
    \centering
    \includegraphics[width=\columnwidth,clip=]{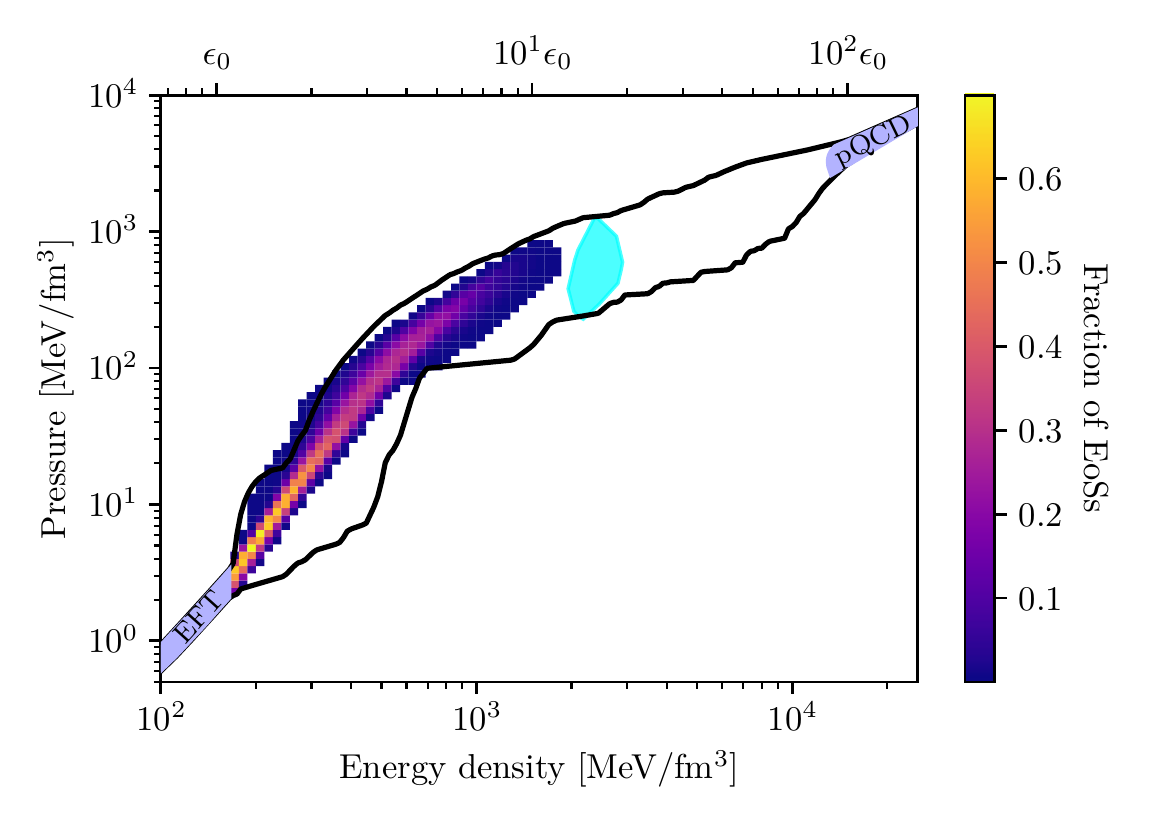} \\
    \includegraphics[width=1.02\columnwidth,clip=]{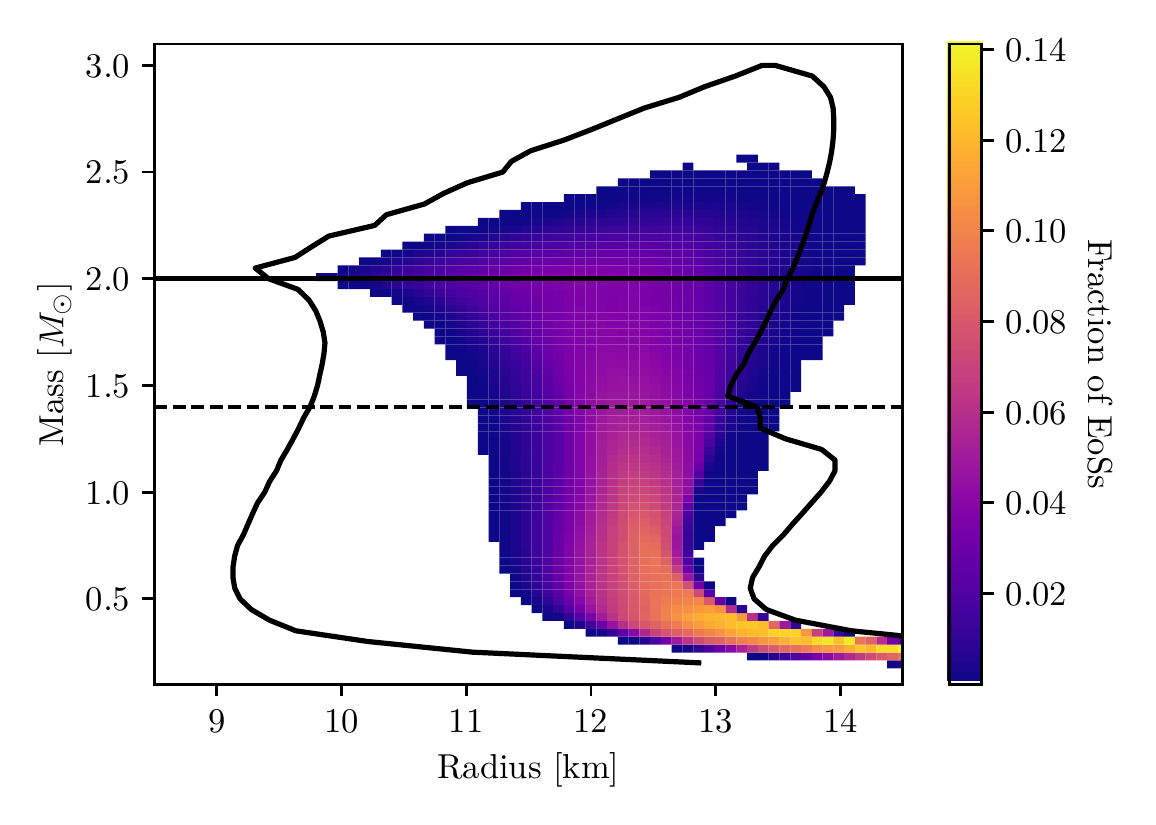}
    \caption{Regions of allowed pressures $p$ and central energy densities $\epsilon$ (top panel) and the corresponding $M$--$R$ region (bottom panel) along stable NS sequences for all \eoss\ within our ensemble, after imposing the two astrophysical constraints $M_{\rm TOV} > 2 M_\odot$ and $\tilde{\Lambda}_{\rm GW170817} < 720$. In the top panel, the cyan regions corresponds to our matching $p$--$\epsilon$ region at $n=10\nsat$, while in the bottom panel, the black outline displays the allowed region of \eoss\ from \citet{Annala:2021gom}, which does not include explicit PTs.}
    \label{fig:ep_MR_PRX}
\end{figure}

\section{Effects of PTs on mass and radius}
\label{sec:MR_discussion}

\begin{figure}[t]
    \centering
    \includegraphics[width=\columnwidth,clip=]{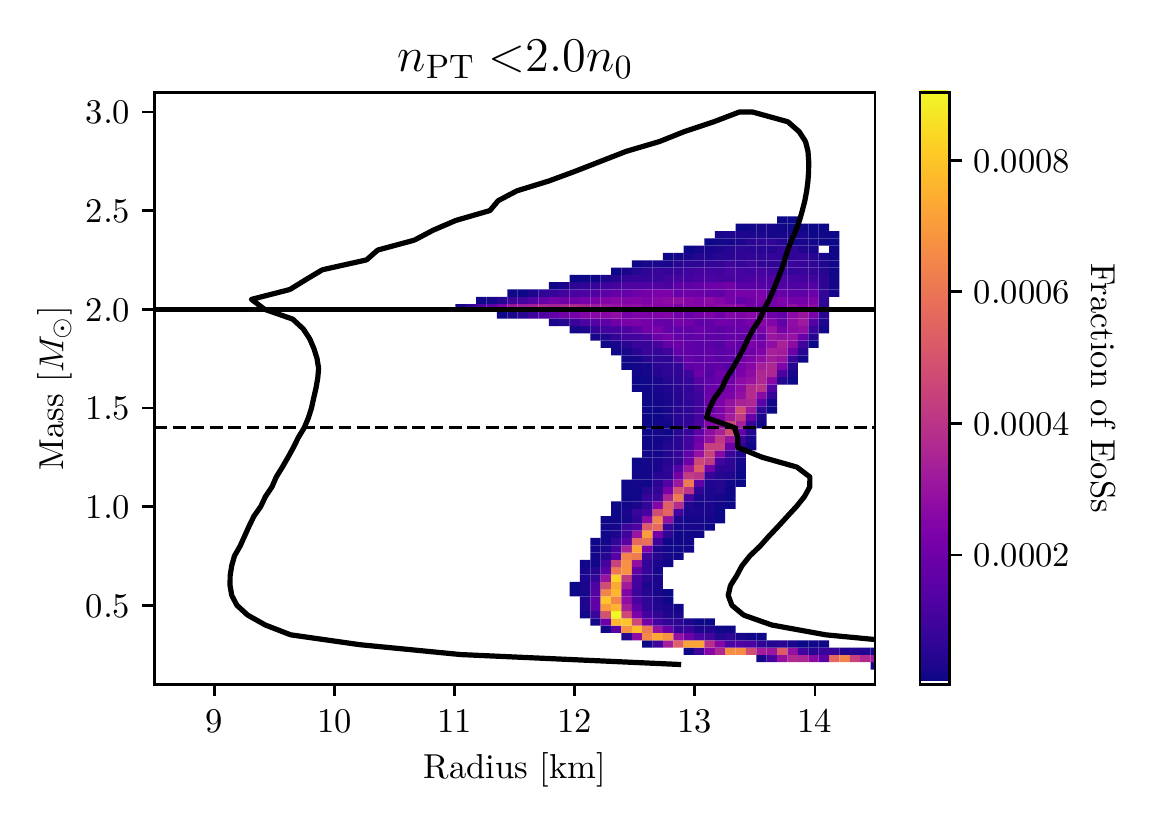} \\
    \includegraphics[width=\columnwidth,clip=]{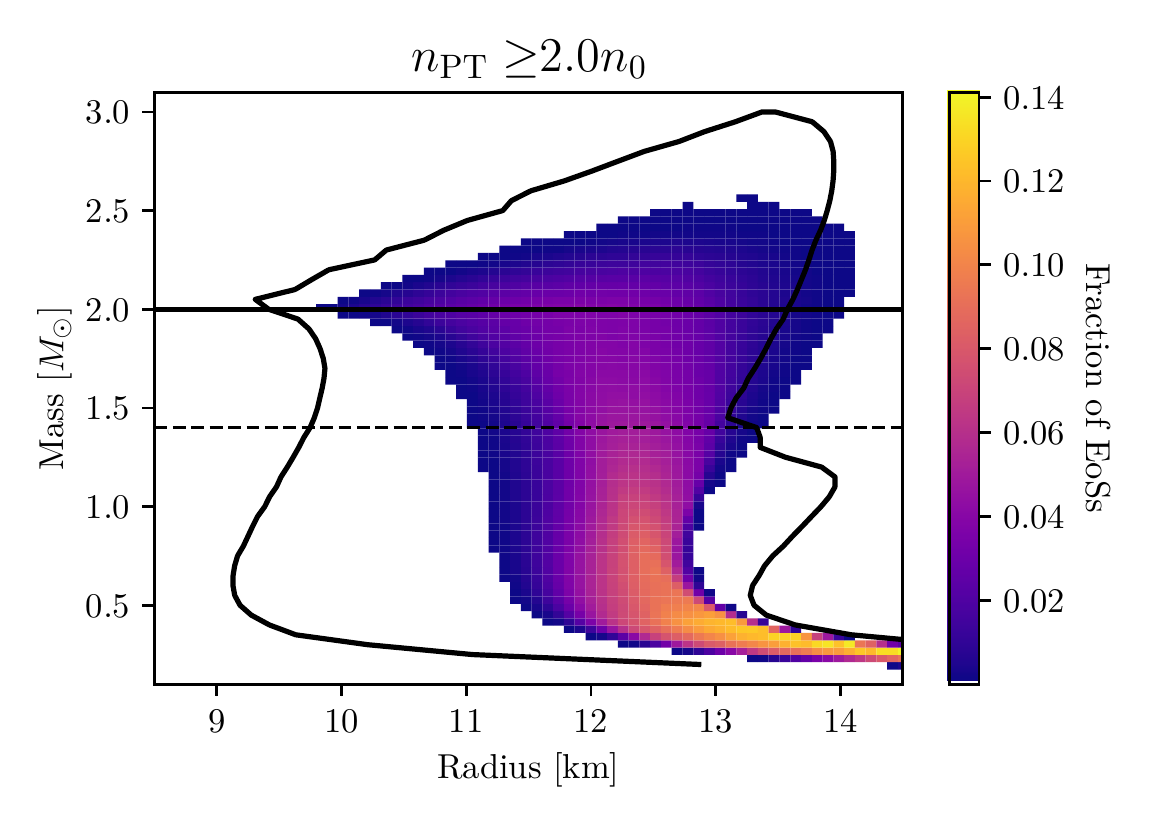}
    \caption{Same $M$--$R$ region as in the bottom panel of \cref{fig:ep_MR_PRX} but this time keeping only \eoss\ with transition densities ${\nPT < 2.0 \nsat}$ (top panel) or ${\nPT \geq 2.0 \nsat}$ (bottom panel). Note the different scales in the colorbars.}
    \label{fig:mr_split_on_nPT_2nsat}
\end{figure}

In \cref{fig:ep_MR_PRX}, we compare the \eos\ and $M$--$R$ regions obtained from our ensemble with first-order PTs matched to the \citet{Annala:2021gom} region at $n=10 \nsat$ to the allowed regions (black lines) obtained in that work. This comparison has the advantage that we can directly see the impact of strong PTs. From the results of \cref{fig:ep_MR_PRX} we observe that PTs extend the range of allowed NS radii consistent with current astrophysical observations beyond the $13.5$~km found in \citet{Annala:2021gom}. A closer inspection into those \eoss\ that feature the largest radii at masses around or above $1.4M_\odot$ reveals that they originate from \eoss\ featuring strong PTs beginning at low densities. This is explicitly seen in \cref{fig:mr_split_on_nPT_2nsat} where we divide the results into \eoss\ with transition densities ${\nPT < 2.0 \nsat}$ or ${\nPT \geq 2.0 \nsat}$, demonstrating that the largest-radius solutions primarily stem from early PTs below $2 \nsat$ NSs. Moreover, the early-PT \eoss\ are found to primarily feature values of the parameter $s \lesssim 0.2$, and hence correspond to matching to the softer part of the chiral EFT band of \citet{Hebeler:2013nza}. These \eoss\ then rapidly stiffen prior to the PT, which is necessary to fulfil the astrophysical constraints, in particular the existence of $2 M_\odot$ NSs. For the early PTs the majority of the \eoss\ with $s > 0.2$, arising from the harder part of the chiral EFT band, fail to pass the tidal-deformalibity constraint.

\begin{figure}[t]
    \centering
    \includegraphics[width=0.875\columnwidth,clip=]{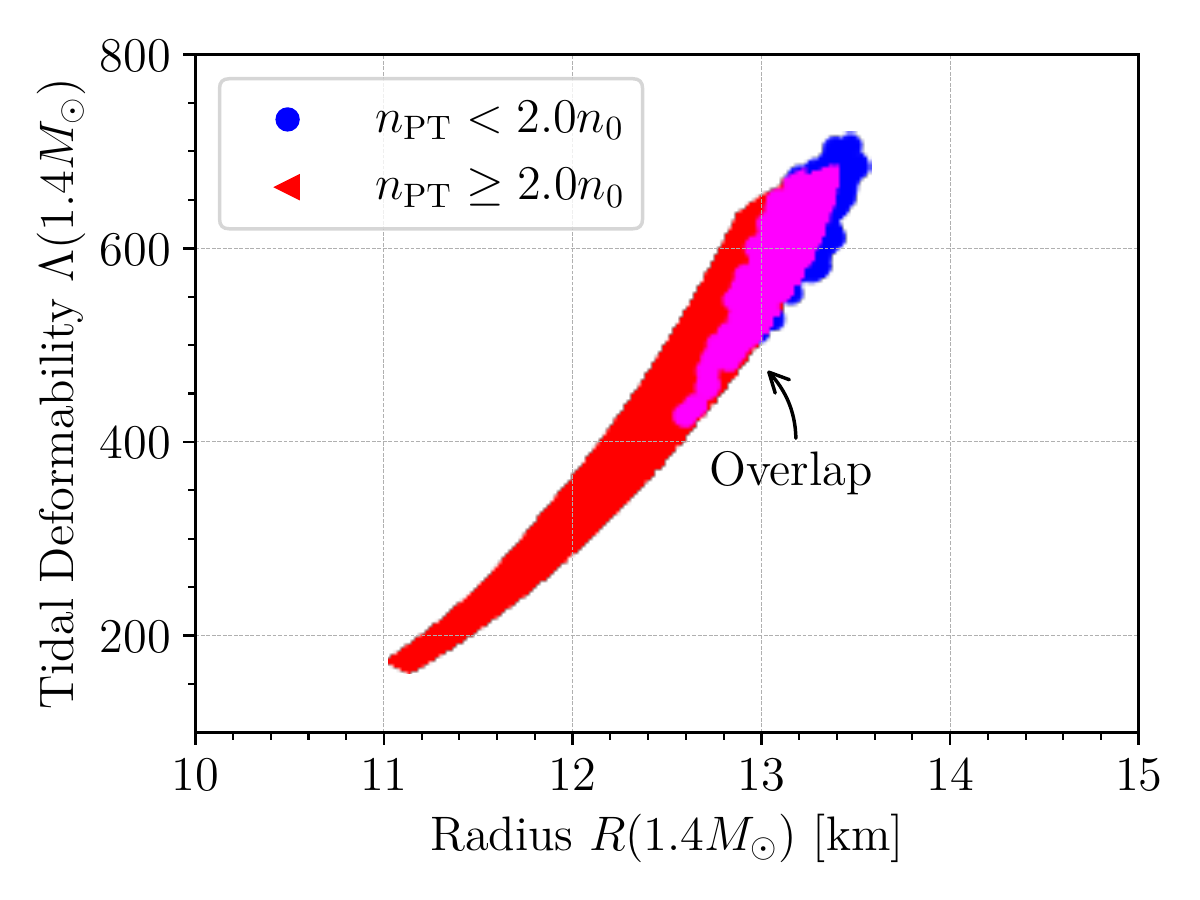} \\
    \includegraphics[width=\columnwidth,clip=]{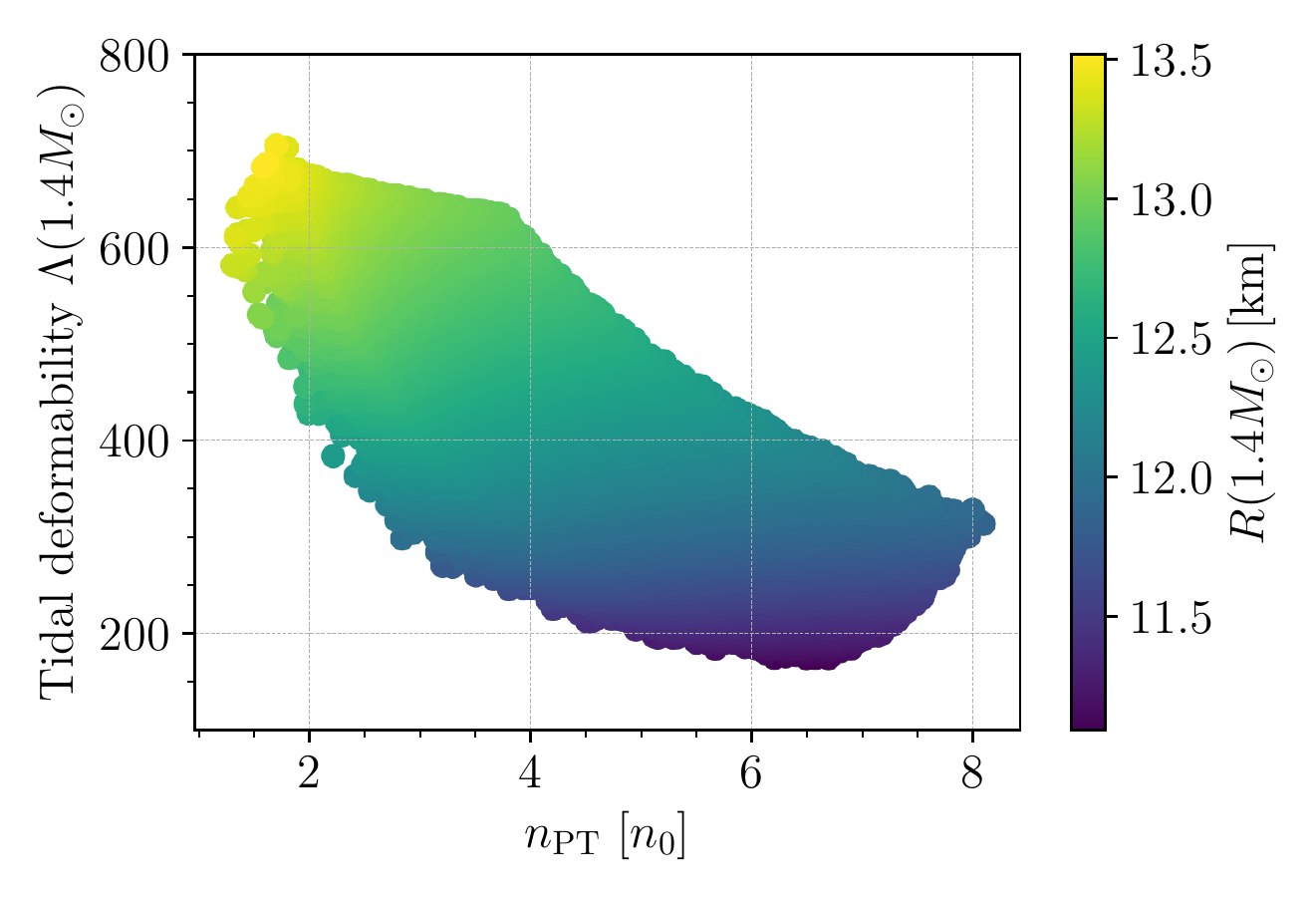}
    \caption{Top panel: Correlation of  $R(1.4 M_\odot)$ and $\Lambda(1.4M_\odot)$, divided into \eoss\ with transition densities $\nPT < 2 \nsat$ (blue) and $\nPT \geq 2 \nsat$ (red). The overlap of these regions is shown in blended color. Bottom panel: Correlation of $\Lambda(1.4M_\odot)$ with the transition density $n_\mathrm{PT}$, with the points colored by their $R(1.4M_\odot)$ values. Note that the largest-radius stars exhibit early PTs.}
    \label{fig:r_lam_nPT_correlations}
\end{figure}

The occurrence of these larger-radius stars can also be traced to the broadening of the correlation between the radius and tidal deformability of a $1.4 M_\odot$ star. This is shown in the top panel of \cref{fig:r_lam_nPT_correlations}, again grouped into early PTs, ${\nPT < 2.0 \nsat}$, and ones occurring later, ${\nPT \geq 2.0 \nsat}$. Phase transitions are seen to broaden the correlation by allowing stellar configurations built with early-PT \eoss\ to reach radii beyond $13.5$~km, confirming the observations of \citet{Han:2018mtj}. 
The bottom panel of \cref{fig:r_lam_nPT_correlations} further demonstrates the linking of early PTs with larger radii. In this figure, we also observe an anticorrelation between $R(1.4M_\odot)$ and $\nPT$, as was previously found in \citet{Miao:2020yjk}. Finally, we note that due to the necessary fine-tuning to fulfill all constraints, these early PTs form a very small fraction of our \eos\ ensemble, so that these large-radius stars are expected to obtain small posterior distributions in a corresponding Bayesian analysis.

\section{Matching to general $\n = 10\nsat$ region}
\label{sec:general_matching}

\begin{figure}[t]
    \centering
    \includegraphics[width=\columnwidth,clip=]{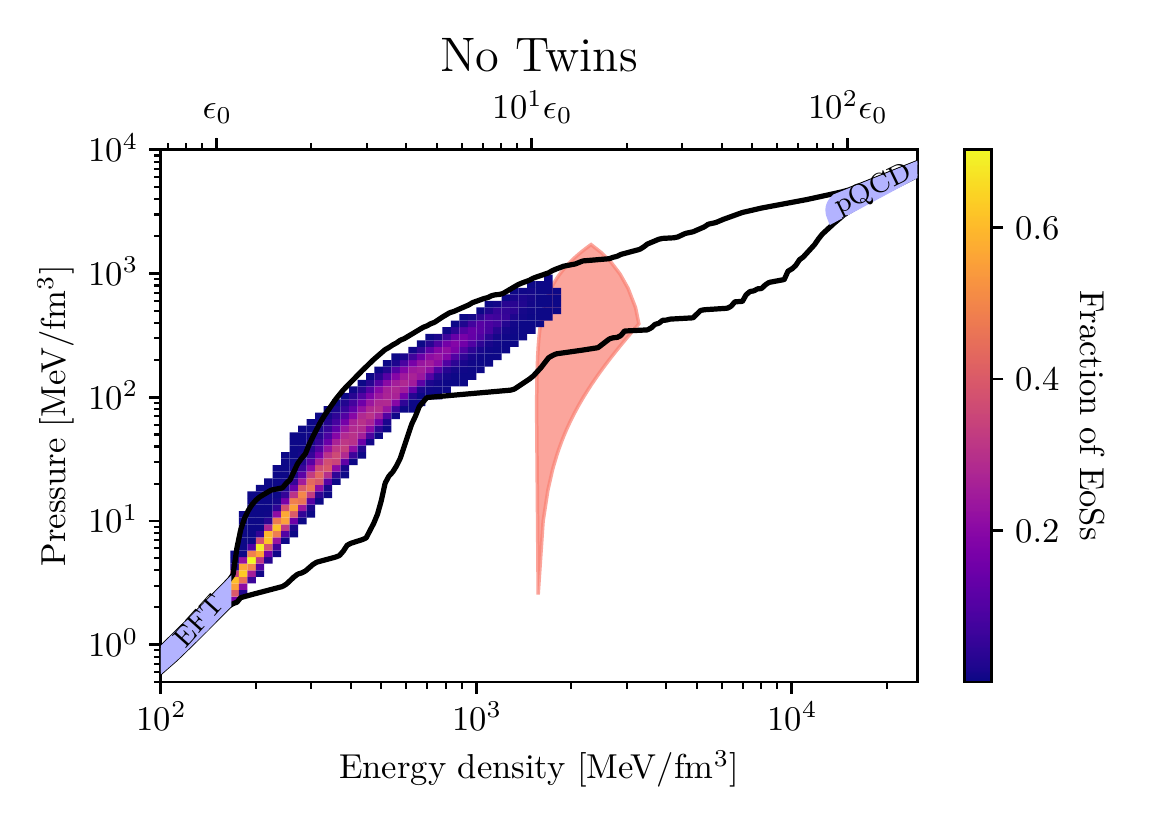} \\
    \includegraphics[width=0.85\columnwidth,clip=]{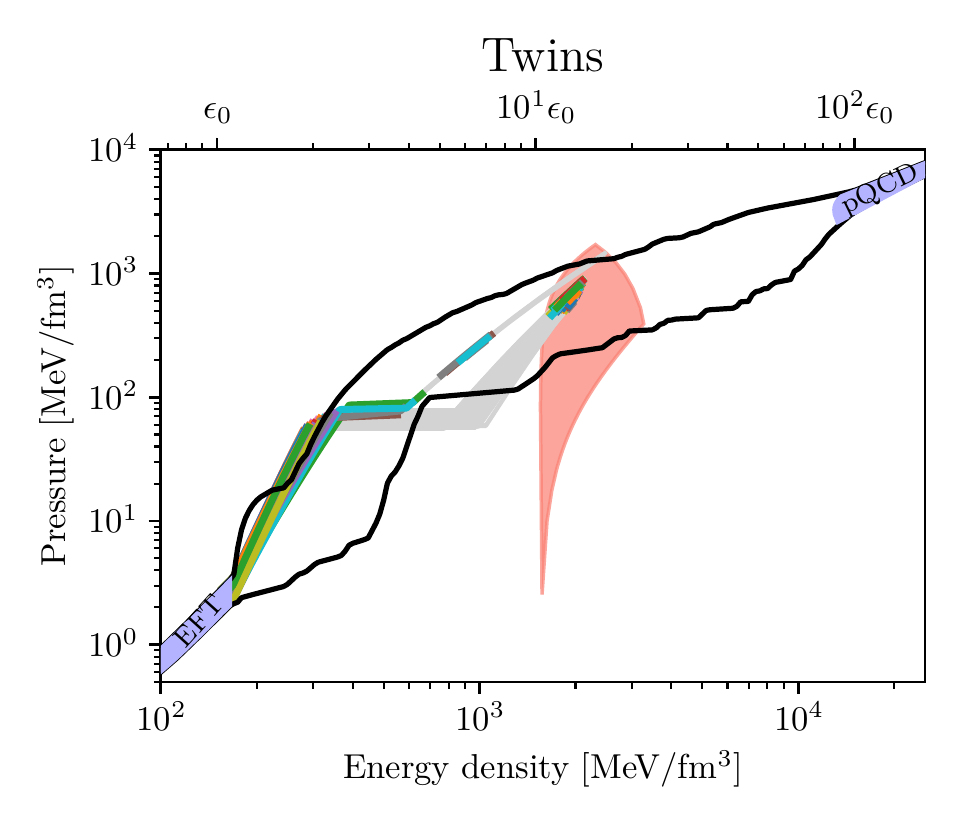}
    \caption{Top panel: Region of allowed central $\epsilon$ and $p$ points along the stable NS sequence with twin-star configurations removed, when using the more general $n=10 \nsat$ matching region of \citet{Komoltsev:2021jzg} (shown in salmon color), after imposing the two astrophysical constraints $M_{\rm TOV} > 2 M_\odot$ and $\tilde{\Lambda}_{\rm GW170817} < 720$. Bottom panel: Individual $p$--$\epsilon$ sequences corresponding to the twin-star solutions, with the unstable branches shown in light gray.}
    \label{fig:ep_KoKu_stab_points}
\end{figure}

We now turn to the more general \eos\ ensemble matched to the maximal $n=10 \nsat$ region of \citet{Komoltsev:2021jzg}. In the top panel of \cref{fig:ep_KoKu_stab_points}, we show the allowed region of $\epsilon$ and $p$ probed within stable NSs using this construction. For comparison, we also show the boundary from the analysis of \citet{Annala:2021gom} (black lines) for the entire NS \eos. Similar to the $M$--$R$ region discussed above, we observe that when PTs are included, the set of allowed \eos~points extends beyond the previous $p$--$\epsilon$ region towards larger pressure values at low and moderate energy densities. 
A similar increase in the $p$--$\epsilon$ region is also seen in \cref{fig:ep_MR_PRX}. 
In particular, \eoss\ that are very stiff at low densities exit the allowed region from \citet{Annala:2021gom} but are still permitted within our present construction. These are precisely the same types of \eoss\ that lead to the large-radius stars in Sec.~\ref{sec:MR_discussion}. In addition, in our more general ensemble we find a small number of \eoss\ with twin-star (or third family) solutions \citep{Gerlach:1968zz,Gerlach:thesis,Schertler:2000xq}, in which the stable stellar sequence first becomes unstable just after the PT, but then becomes stable again at higher central densities following a stiffening of the \eos\ beyond the PT. These individual solutions are shown separately in the bottom panel of \cref{fig:ep_KoKu_stab_points}, and are discussed in detail in Sec.~\ref{sec:twins}.

\begin{figure}[t]
    \centering
    \includegraphics[width=1.05\columnwidth,clip=]{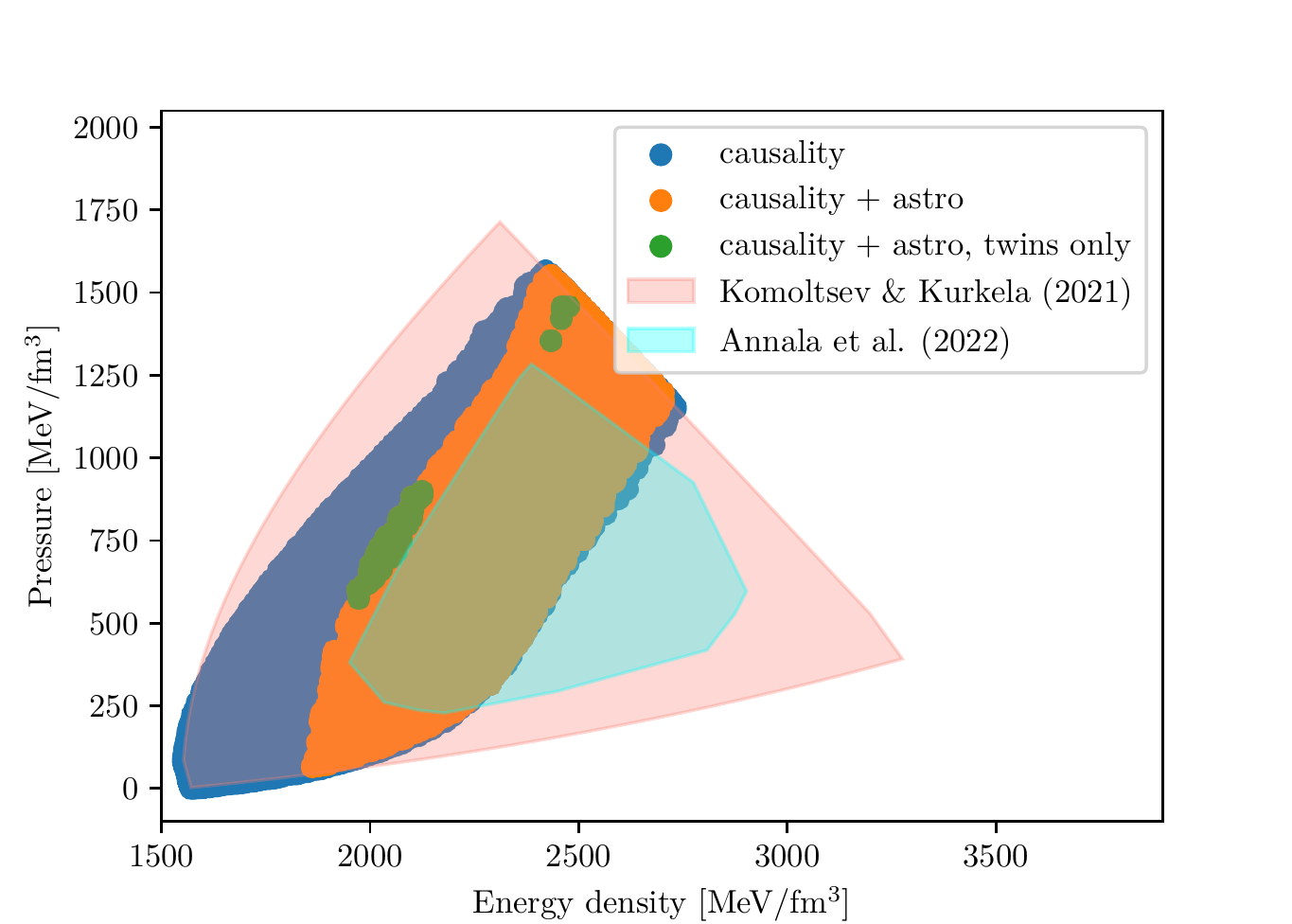}
    \caption{Range of allowed $\epsilon$ and $p$ values at $10\nsat$. The cyan region corresponds to that shown in \cref{fig:ep_MR_PRX}, while the salmon one is the more general $n = 10 \nsat$ matching region of \citet{Komoltsev:2021jzg}. The points denoted by ``causality'' correspond to our construction prior to applying astrophysical constraints.}
    \label{fig:ep_KoKu_10n0_matching_reg}
\end{figure}

To more closely examine the differences between our current construction and previous results, we study the $p$--$\epsilon$ matching region at $n=10\nsat$  in \cref{fig:ep_KoKu_10n0_matching_reg}. We show the regions from \citet{Annala:2021gom} and \citet{Komoltsev:2021jzg}, with the points from our current ensemble overlayed. The \citet{Annala:2021gom} region indeed is a subset of the maximal region, as it must be. We also observe that before applying the astrophysical constraints, our current simple construction (denoted by ``causality'' in \cref{fig:ep_KoKu_10n0_matching_reg}) covers about half the area of the two regions (either \citet{Annala:2021gom} or \citet{Komoltsev:2021jzg}), with a significant number of \eoss\ lying outside the \citet{Annala:2021gom} region. More interestingly, even after applying the astrophysical constraints, we still find a significant number of \eoss\ that lie outside of that region, which emphasizes the effect of including PTs in a general \eos\ ensemble. We also note that the impact of the pQCD constraint from \citet{Komoltsev:2021jzg} is visible in this figure as the cut to the upper right of the causal and astrophysically allowed regions.

\begin{figure}[t]
    \centering
    \includegraphics[width=0.9\columnwidth,clip=]{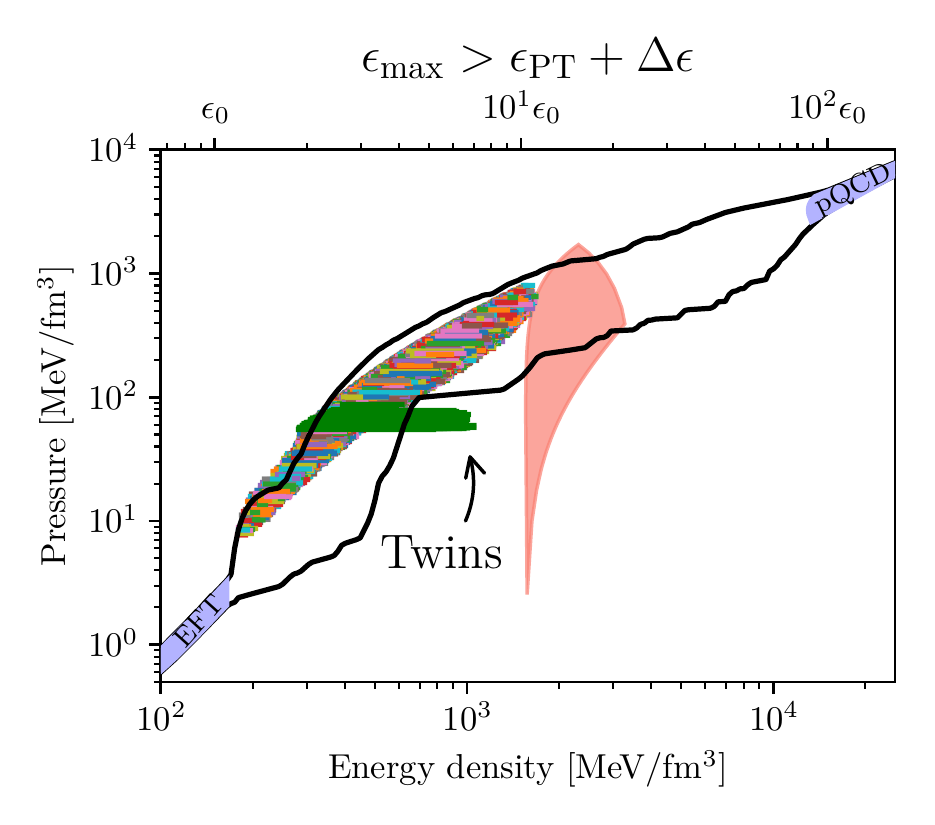} \\
    \includegraphics[width=0.9\columnwidth,clip=]{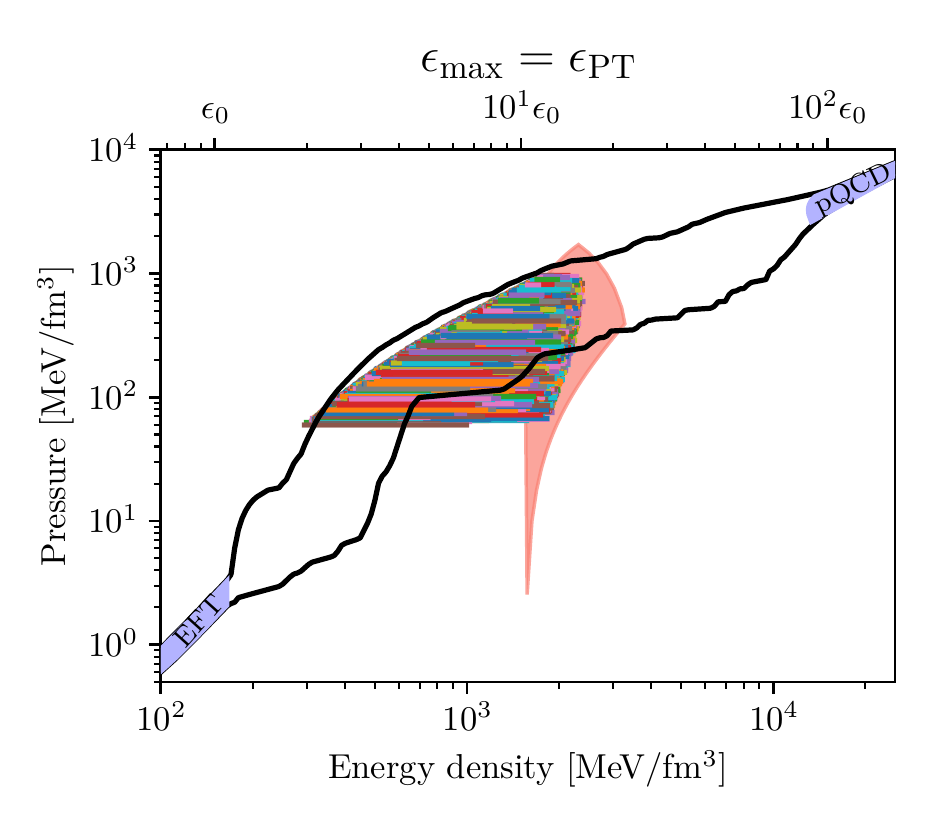}
    \caption{Locations of the PTs (each drawn as a line with random color) in our ensemble matched to the maximal $n=10 \nsat$ region from \citet{Komoltsev:2021jzg} shown in salmon color. Top panel: \eoss\ for which the PT completes within the stable NS sequence, with twin solutions shown in green. Note that for most of the twin solutions, the central densities occur at much higher densities than the ends of the PTs. Bottom panel: \eoss\ for which the PT destabilizes the NS sequence.}
    \label{fig:ep_with_PT_lines}
\end{figure}

In \cref{fig:ep_with_PT_lines}, we show the $p$--$\epsilon$ region with the individual PTs from our ensemble, distinguished by whether any point beyond the PT is stable (top panel: $\epsilon_\mathrm{max} > \ePT + \De$) or whether the PT terminates the NS sequence (bottom panel: $\epsilon_\mathrm{max} = \ePT$), with $\epsilon_\mathrm{max}$ denoting the highest energy density reached along the stable stellar sequence.
Practically, we define this $\epsilon_\mathrm{max}$ as the last central density whose forward finite-difference mass derivative satisfies $dM / dn_\mathrm{central} > 0$, using a high-resolution grid in number density. We note that our resulting definitions of stability or destabilization are not strictly equivalent to the microscopic stability analysis conducted in \citet{Seidov_PT_destab, Schaeffer_PT_destab, Lindblom:1998dp}. As discussed in \cite{Lindblom:1998dp}, such microscopic stability does not necessarily lead to macroscopic cores of high-density matter beyond the PT, and hence our definition will neglect some cores of negligible size. Also in this context, we note that the above classification into PTs which do or do not terminate the stellar sequence does not represent a clean delineation within the full ensemble since there are \eoss\ with arbitrarily small values of $\Gamma_2$ that effectively prolong the PT beyond the value of the $\Dn$ parameter. Within these and future figures, when presenting results for \eoss\ that possess stable stellar configurations beyond the PT, we additionally remove any \eoss\ possessing $\Gamma_2 < 0.15$ from the ensemble.\footnote{This removes less than 2\% of the ensemble.}
We clearly observe a minimal $\ePT \approx 275$~MeV/fm$^3$ for PTs that terminate the NS sequence, which is set by the astrophysical $2 M_\odot$ constraint. We also see that \eoss\ exhibiting twin-star solutions have a particular range of PT parameters, which will be further studied below. Moreover, \cref{fig:ep_with_PT_lines} shows how the PTs in our general ensemble go beyond the region from \citet{Annala:2021gom} without explicit PTs.

\begin{figure}[t]
    \centering
    \includegraphics[width=\columnwidth,clip=]{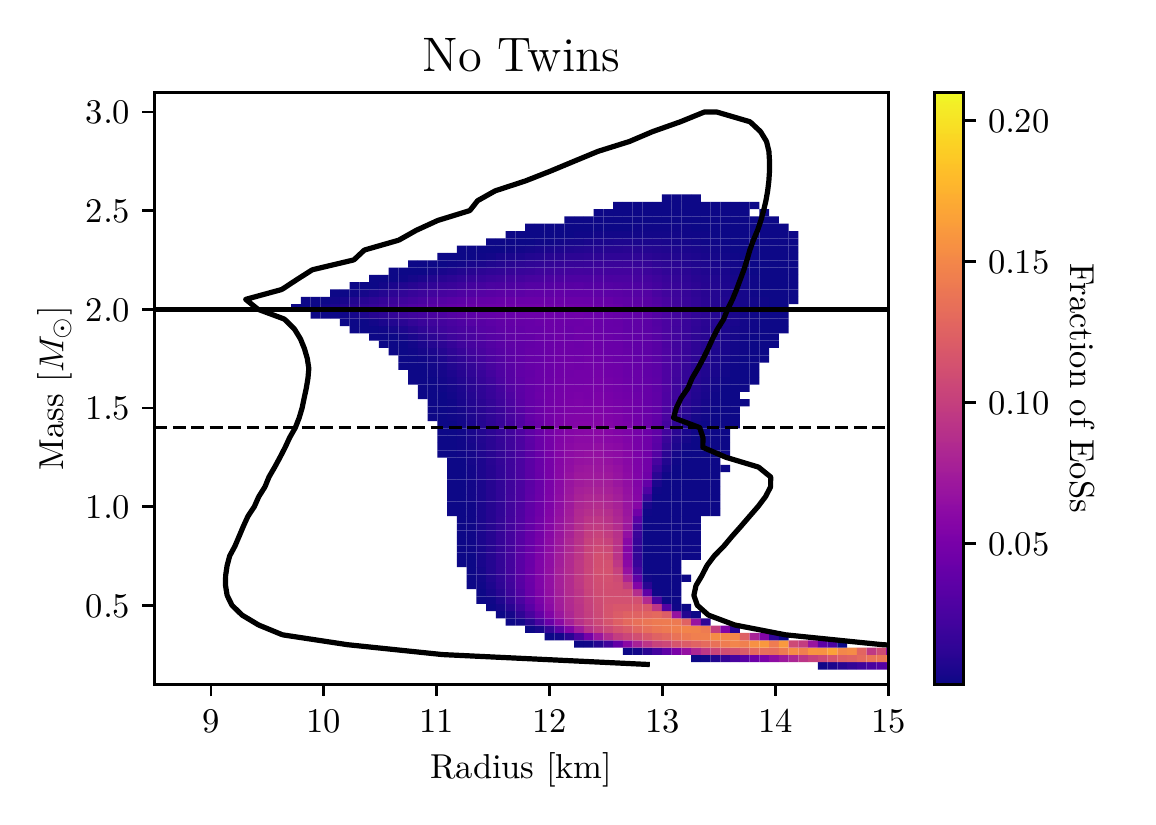} \\
    \includegraphics[width=0.875\columnwidth,clip=]{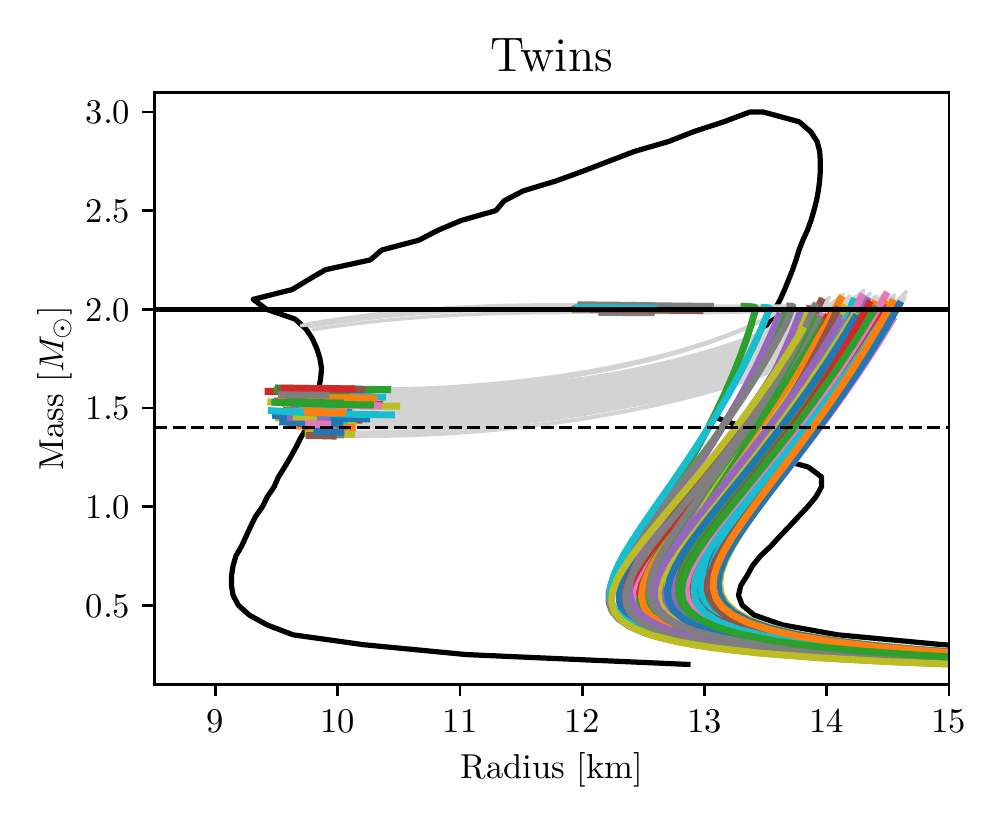}
    \caption{Top panel: Region of allowed masses and radii when matching to the \citet{Komoltsev:2021jzg} region at $n = 10\nsat$, with twin-star configurations removed. Bottom panel: $M$--$R$ sequences corresponding to the individual twin-star solutions, with the unstable branches shown in light gray.} 
    \label{fig:mr_koku}
\end{figure}

Turning then to the $M$--$R$ sequences in \cref{fig:mr_koku}, we find the $M$--$R$ region without twin solutions to be very similar to \cref{fig:ep_MR_PRX} above, following from matching to the high-density region from \citet{Annala:2021gom}. This is not surprising, since the constraints at small and large $R$ follow directly from the astrophysical $2 M_\odot$ and GW170817 constraints, respectively.
The main effect of the more general $10 \nsat$ matching region is to allow for twin solutions with PTs, which are plotted in the bottom panel of \cref{fig:mr_koku}.

\begin{figure}[t]
    \centering
    \includegraphics[width=\columnwidth,clip=]{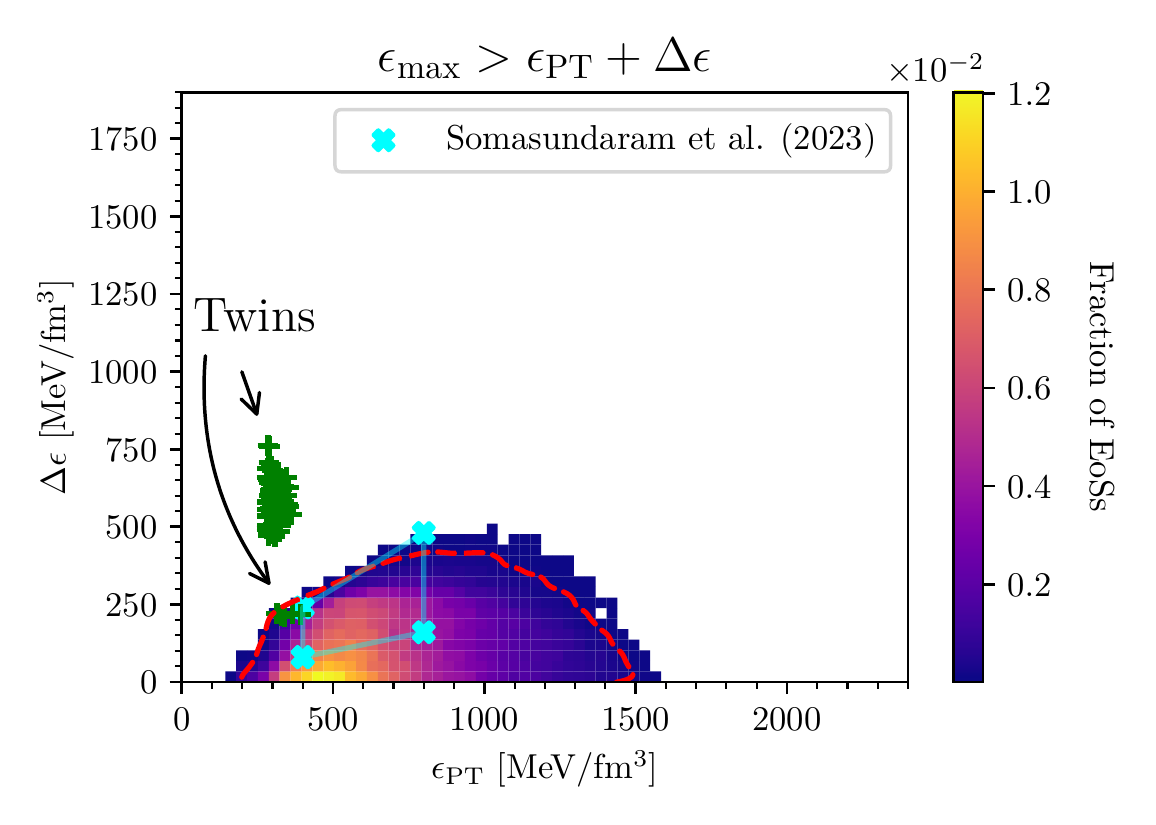} \\
    \includegraphics[width=\columnwidth,clip=]{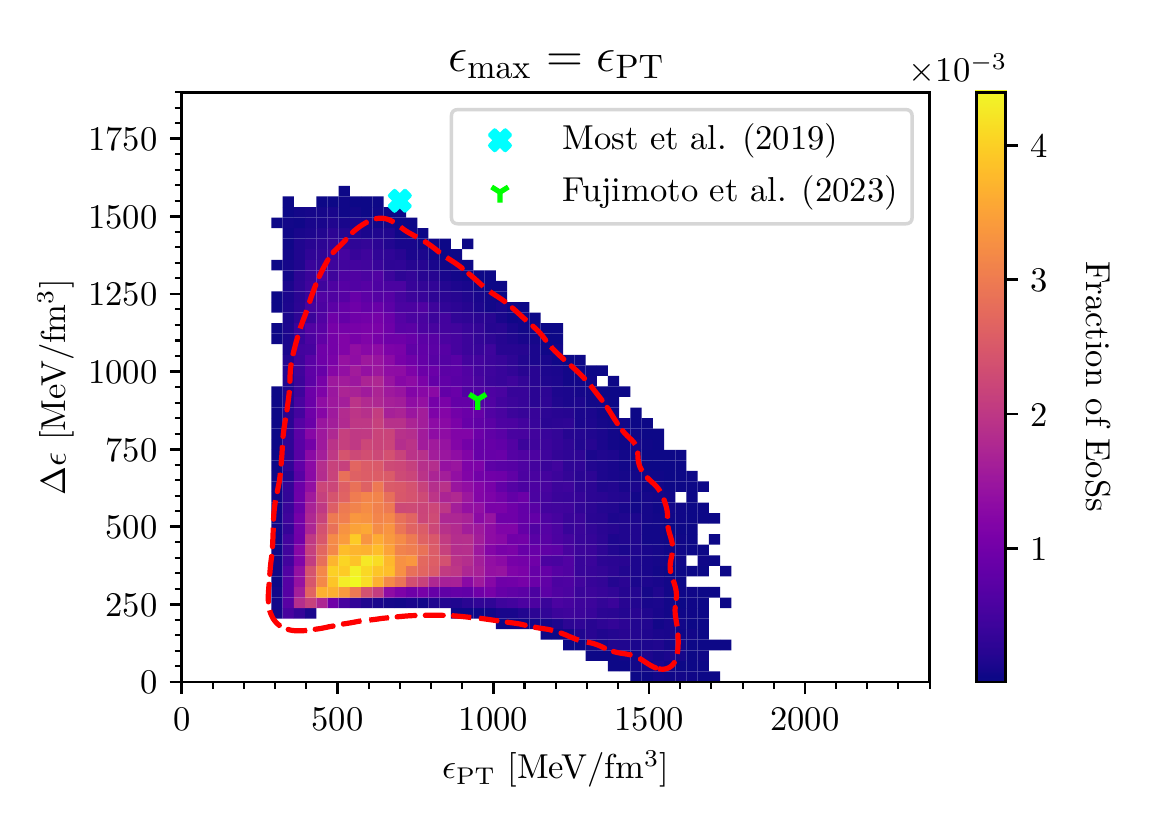}
    \caption{Allowed values of $\De$ and $\ePT$ for PTs that either complete in stable NSs (top panel) or destabilize the NS sequence (bottom panel) when matching to the \citet{Komoltsev:2021jzg} region at $n = 10\nsat$. The green pluses denote the twin solutions, and the remaining markers show the parameters corresponding to \eoss\ studied in the literature (see \cref{tab:PT_literature}). The dashed red line denotes the 99.5\% confidence contour when matching to the \citet{Annala:2021gom} region at $n=10 \nsat$.}
    \label{fig:e_PT_properties}
\end{figure}

\begin{table}[t]
\centering
\addtolength{\tabcolsep}{-4pt}
\fontsize{7.8pt}{10pt}\selectfont
\begin{tabular}{lll} \toprule
Reference & PT start & PT size \\ \midrule
\citet{Bauswein:2018bma} & 2.94--3.63$\,\nsat$ & 0.19--0.75$\,\nsat$ \\
\citet{Han:2018mtj} & 1--3.5$\,\nsat$ & 0.17--3.11$\,\nsat$ \\
\midrule
\citet{Most:2018eaw} & 700\,MeV/fm$^3$ & 1550\,MeV/fm$^3$\\
\citet{Somasundaram:2021clp} & 400--800\,MeV/fm$^3$ & 80--480\,MeV/fm$^3$ \\
\citet{Fujimoto:2022xhv} & 950\,MeV/fm$^3$ & 910\,MeV/fm$^3$ \\
\bottomrule
\end{tabular}
\caption{Examples of PTs studied in the literature, with the upper two reported in baryon densities and the lower three in energy densities.}
\label{tab:PT_literature}
\end{table}

\begin{figure}[t]
    \centering
    \includegraphics[width=\columnwidth,clip=]{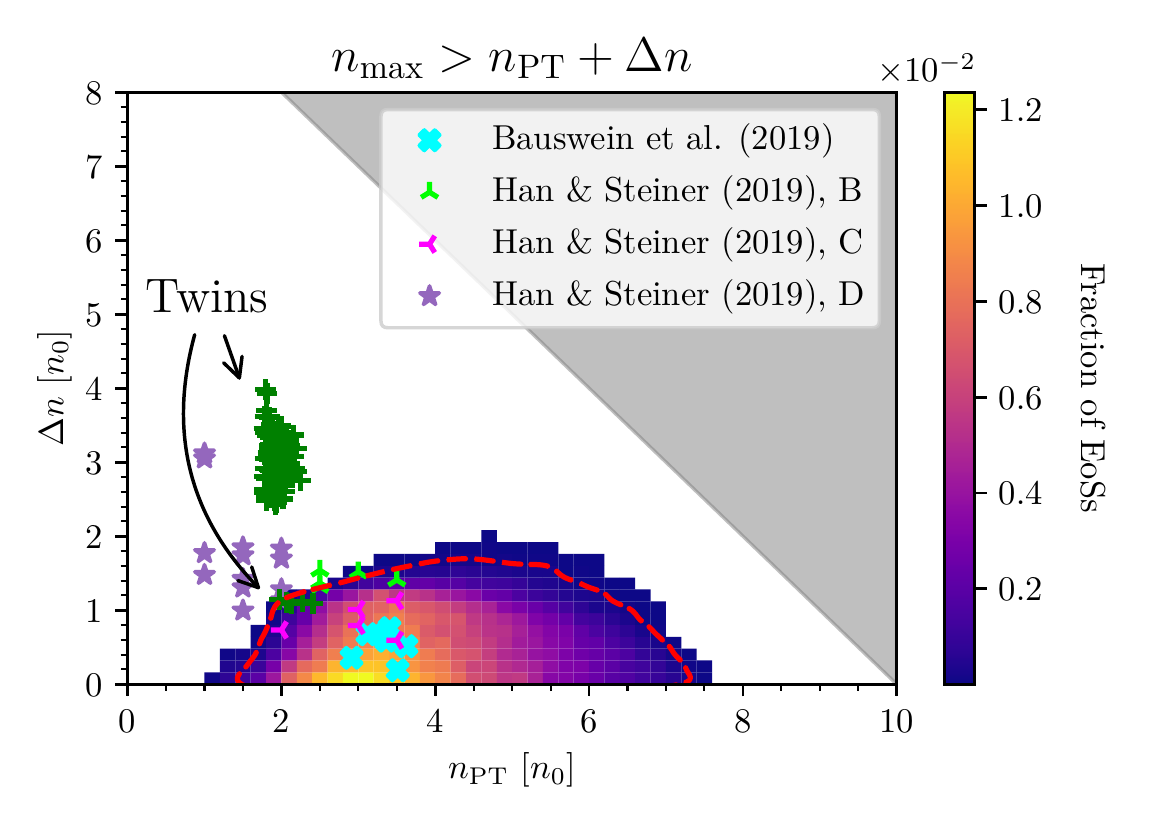} \\
    \includegraphics[width=\columnwidth,clip=]{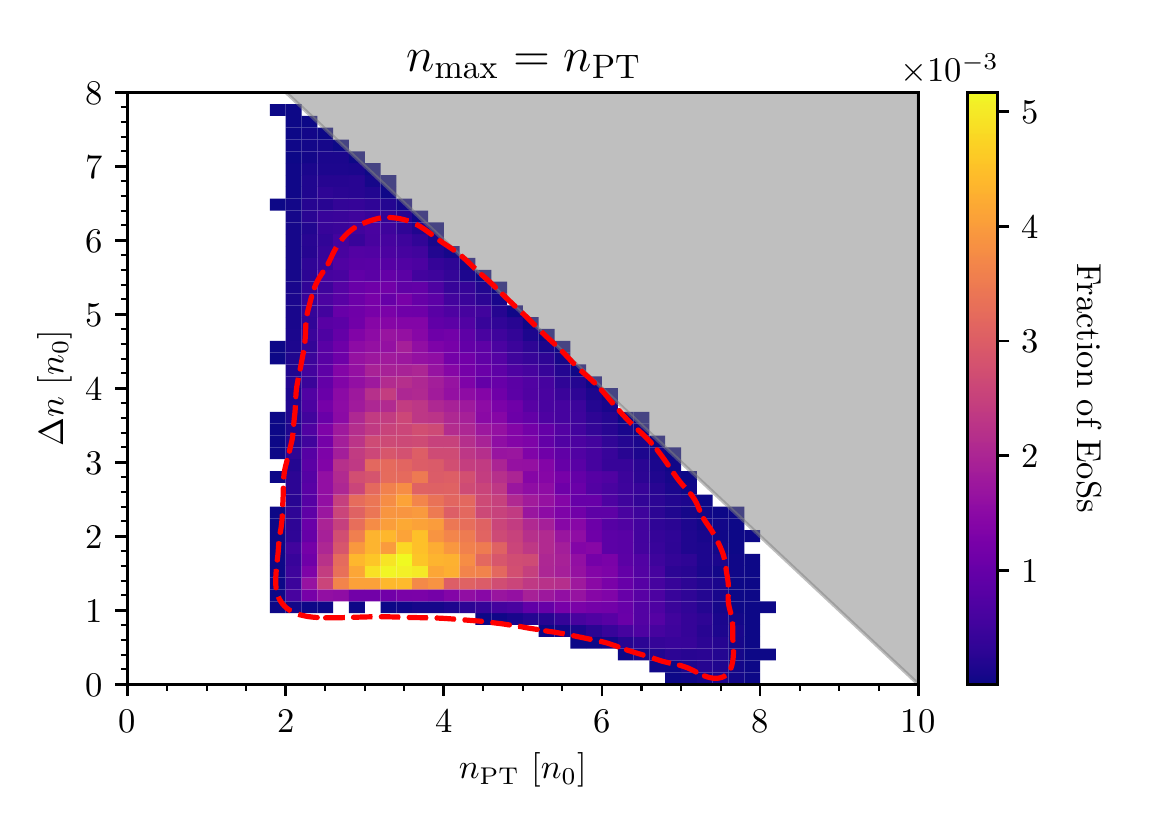}
    \caption{Same as \cref{fig:e_PT_properties} but for $\Dn$ and $\nPT$. The gray region denotes the range of parameters that cannot be reached or constrained by our model, due to our choice of performing the high-density matching at $n = 10\nsat$.}
    \label{fig:n_PT_properties}
\end{figure}

Next, we examine the range of allowed PT parameters $\ePT$ and $\De$, or $\nPT$ and $\Dn$, for the \eoss\ we have constructed. To make this discussion more concrete, we also compare our allowed ranges to values that have been studied in the literature, see \cref{tab:PT_literature}. We show in \cref{fig:e_PT_properties} the allowed values of $\ePT$ and $\De$ for PTs that either complete within stable NSs or destabilize the NS sequence, while the corresponding values of $\nPT$ and $\Dn$ are shown in \cref{fig:n_PT_properties}. In both figures, we separate the twin solutions from the remaining distribution of allowed ranges, and for comparison show the 99.5\% confidence contour when matching to the \citet{Annala:2021gom} region at $n=10 \nsat$ (red dashed lines). We emphasize that, due to our choice of performing the high-density matching at $n = 10\nsat$, there is a range of parameters that cannot be reached in our simple model,  indicated as the grey region in 
\cref{fig:n_PT_properties}. Likewise there is a similar region in \cref{fig:e_PT_properties}, but its boundary is not easy to parameterize.

Our first observation from these figures is that there is a substantial overlap in the PT parameters in the $\De$--$\ePT$ or $\Dn$--$\nPT$ planes that can either destabilize the sequence or complete within stable NSs, leading to high-mass cores of matter beyond the phase transition. We will discuss these cores in more detail below in Sec.~\ref{sec:cores}. Next, we compare our results to PTs studied in the literature, collected in \cref{tab:PT_literature}. We mark these points in Figs.~\ref{fig:e_PT_properties} and~\ref{fig:n_PT_properties}, placing them in panels that indicate how the corresponding \eoss\ were constructed: namely whether they were engineered to complete within the stable NS sequence or to destabilize it. We observe that past literature has mostly focused on parameters corresponding to lower transition densities, and that there remains a large range of allowed PT parameters for $\ePT \gtrsim 800$~MeV/fm$^3$ or $\nPT \gtrsim 4 \nsat$ that has not been explored in merger simulations.
Moreover, stronger (destabilizing) PTs with $\De \approx 750$~MeV/fm$^3$ and $\Dn \approx 3\nsat$ at intermediate densities have also not been probed. 
Additionally, we find a substantial range of destabilizing PTs that have not been studied, but may lead to interesting dynamics in the post-merger phase of a binary-NS merger.

In more detail, the PT scenarios explored in \citet{Bauswein:2018bma} and \citet{Somasundaram:2021clp} overlap nicely with our most likely range of PT parameters, while \citet{Most:2018eaw} studied a particularly strong PT\footnote{Note that the upper-right boundary of our region originates from our high-density matching at $n=10 \nsat$.} and \citet{Fujimoto:2022xhv} utilized one of intermediate strength. However, there remain even earlier and weaker PTs that destabilize the star and are allowed within our construction. Finally, \citet{Han:2018mtj} [see also \citet{Alford:2013aca}] examined three different scenarios (labeled B, C, D) of \eoss\ with $n_\mathrm{max} > \nPT + \Dn$, the parameters of which are shown in the top panel of \cref{fig:n_PT_properties}. Their classes C and D correspond to stellar sequences that are connected or disconnected at the PT, respectively, meaning that the point just beyond the PT at $\nPT + \Dn$ is either stable or unstable. In the latter case (D), the sequence later includes another stable twin-star branch due to a stiffening of the \eos\ beyond the PT. Class B (for ``both'') describes \eoss\ where the point just after the PT is stable, but which later experience a destabilization and then a further stabilization with a twin-star branch. We find that classes B and C, where the point $\nPT + \Dn$ is stable, lie within our region of non-twin-star solutions, while class D lies outside our region, somewhat in the vicinity of our twin-star solutions. 

\section{Twin Stars}
\label{sec:twins}

From the lower panel of \cref{fig:mr_koku} we see that two separate classes of twin-star solutions are generated in our ensemble. One class features a small PT that locally destabilizes the NS sequence, but the sequence is then quickly restabilized by a stiffening of the \eos. These solutions have high-mass twins all possessing $M \approx 2 M_\odot$. For these solutions, the stable stellar sequence is almost connected, and as such, the PT parameters overlap with those of the \eoss\ without twin-star solutions. We note that these very few high-mass-twin \eoss\ lead to $\tilde{\Lambda} \approx 720$, so that they only marginally pass the GW170817 tidal-deformability constraint and would obtain very small posterior distributions in a corresponding Bayesian analysis.

The second and more interesting class of twin-star solutions is more exotic, with a much larger phase transition that destabilizes the stellar sequence for a large range of central densities. These solutions have a low-mass twin branch beginning at a central density close to $n = 10\nsat$ and remaining stable up to the matching density. The \eoss\ featuring low-mass twins have the peculiar feature that the low-density NS branch has large radii and does not on its own pass the GW170817 constraint, but it is the twin-star branch, with masses $M \approx 1.4 M_\odot$, that allows the \eoss\ to pass the gravitational-wave constraint with their extremely small $\Lambda$ values. For these \eoss, the GW170817 event would have been a merger between a twin star and a regular NS, or between two twin stars, as both of these scenarios lead to small binary tidal deformabilities $\tilde\Lambda \ll 720$. We finally remark that in this case it is not clear how these twin stars would form, as they have much smaller gravitational and baryonic masses than the highest-mass NSs along the same \eos. Hence, during their formation from a NS a large amount of material $\sim 0.5M_\odot$ would need to be ejected. In addition, it is unclear whether such twin star solutions would be consistent with the electromagnetic counterparts of GW170817.

We conclude by observing that all of our twin-star solutions feature maximal masses only marginally exceeding $2M_\odot$. While within a more general \eos\ construction, the maximal masses may be somewhat increased, overall these twin-star solutions are not very probable, especially given possible observations of even higher-mass NSs~\citep{Linares:2018ppq,Romani:2022jhd}. 

\section{Cores of high-density matter}
\label{sec:cores}

\begin{figure}[t]
    \centering
    \includegraphics[width=\columnwidth,clip=]{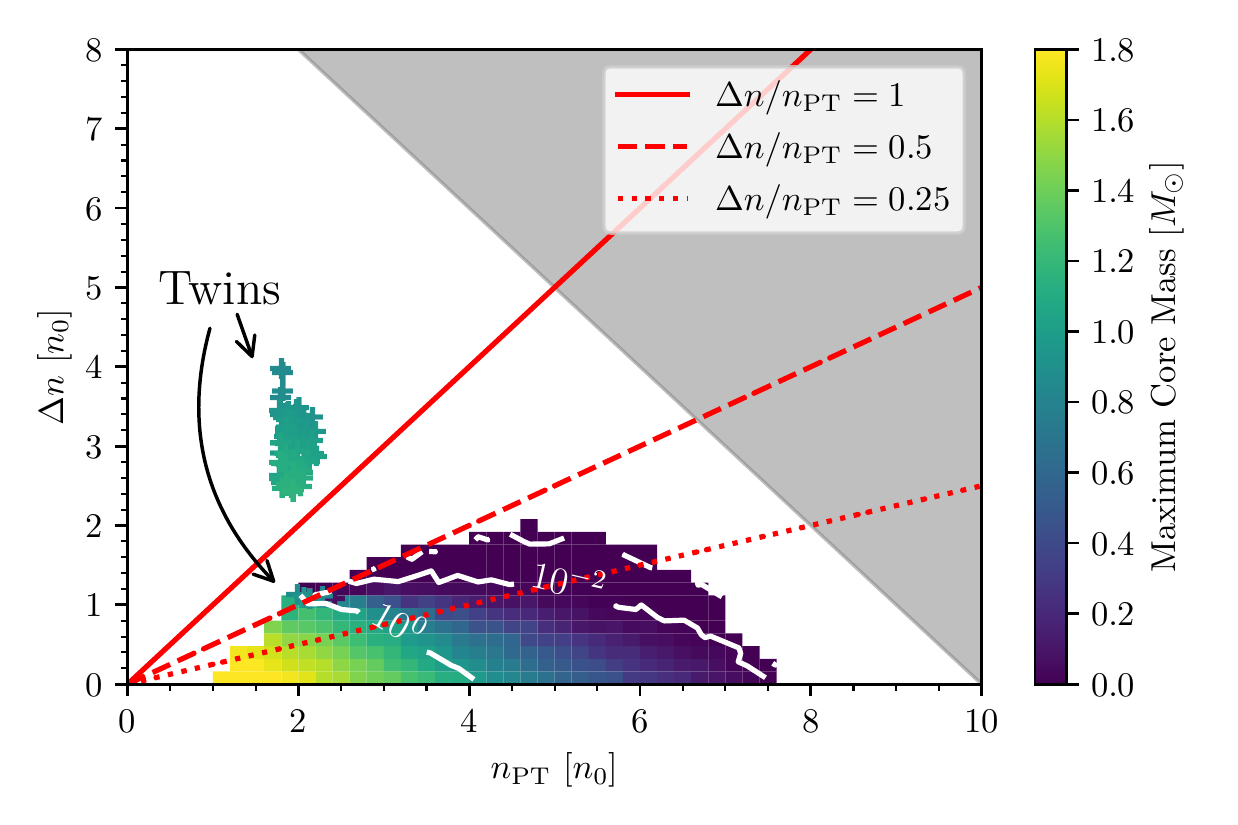}
    \caption{The maximum possible mass
    of the core of high-density matter for \eoss\ with different PT parameters. 
    Also shown are white contour lines of constant core masses (in units of $M_\odot$) to make details of the small-core-mass region more visible. Here, only points leading to nonzero core masses are shown.}
    \label{fig:mr_cores}
\end{figure}

In this section, we explore the cores of high-density matter in stable NSs, which we define as the part of the star described by the second polytrope after the phase transition. We do not speculate on the nature of the high-density phase. 
In \cref{fig:mr_cores} we show the maximum possible mass of the core for given PT parameters. Unsurprisingly, the largest cores are related to transitions with a very early onset, but cores with masses exceeding $0.5M_\odot$ are possible for onsets up to $5\nsat$ (but small $\Dn$).

\begin{figure}[t]
    \centering
    \includegraphics[width=1.03\columnwidth,clip=]{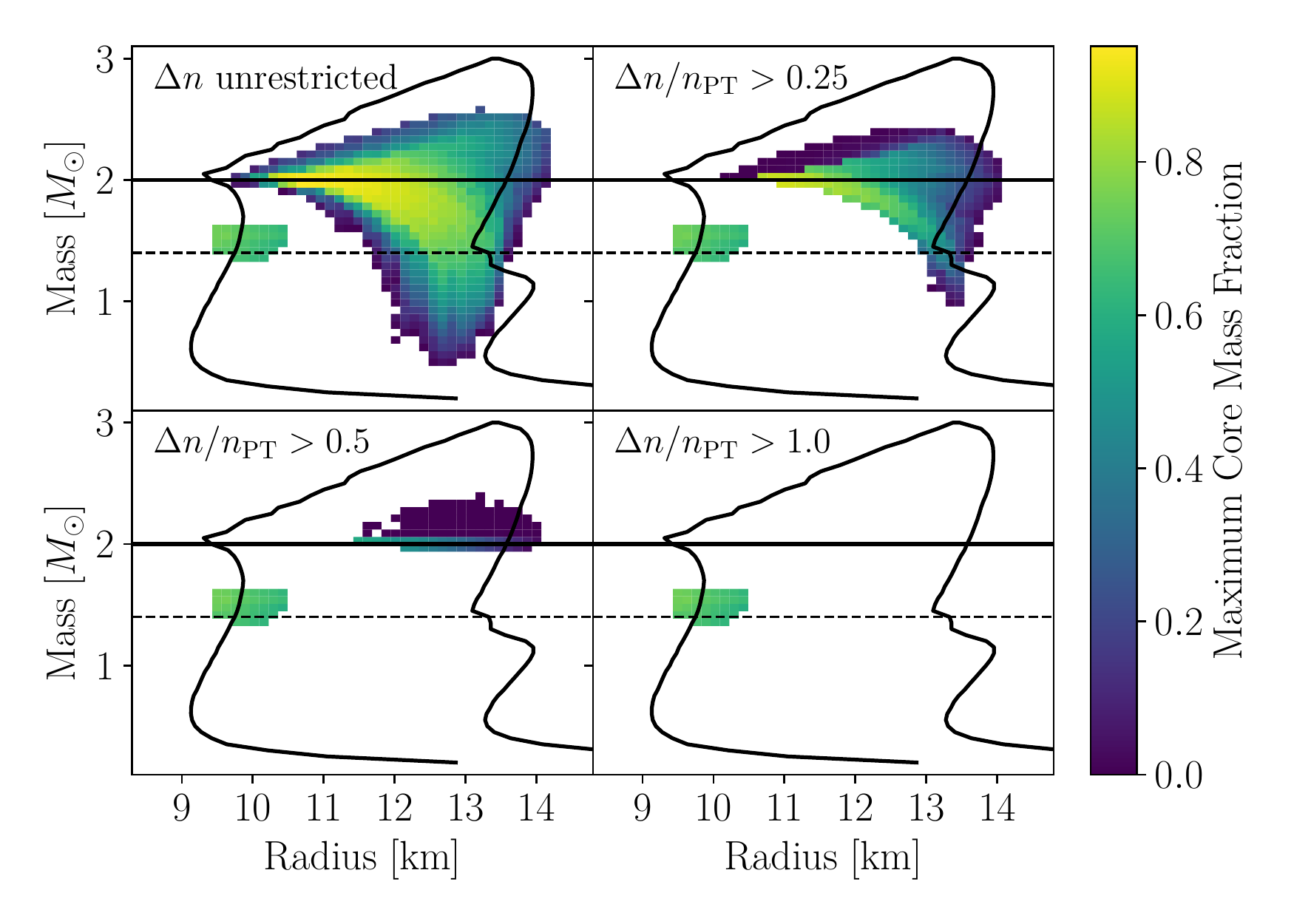}
    \caption{The mass fraction of the core $M_\mathrm{core}/M$ for different cuts on $\Delta n / n$. Note that the cuts shown correspond to the red lines in \cref{fig:mr_cores}.}
    \label{fig:mr_cores_dnOnPT_prog}
\end{figure}

In \cref{fig:mr_cores_dnOnPT_prog}, we display the mass fraction of the core, $M_\mathrm{core}/M$, in the $M$--$R$ diagram, further dissecting the results by the minimal value of $\Delta n/n_{\text{PT}}$ (corresponding to PTs above the corresponding red lines in \cref{fig:mr_cores}). As one can readily observe, excluding the twin-star solutions the largest cores originate from weak phase transitions with $\Delta n/n_{\text{PT}} < 0.5$ and heavier NSs, while the largest PTs lead to small cores at masses $M \gtrsim 2 M_\odot$. This reflects the fact that in our ensemble all sizable PTs begin at fairly high densities, at which point the stars are already quite massive. Note also that the low-mass twin-star configurations possess sizeable cores of high-density matter.

\section{Summary}
\label{sec:summary}

In this work, we have explored bounds for first-order PTs that are compatible with present \eos\ constraints from chiral EFT up to nuclear densities, current NS observations, and pQCD results at very high densities. To this end, we constructed a large ensemble of \eoss\ with a two-polytrope form, always featuring a first-order PT in between, demanding that the lower polytrope connects to the chiral EFT band from \citet{Hebeler:2013nza}, the higher polytrope connects to pQCD constraints at $n=10 \nsat$ from \citet{Annala:2021gom} or \citet{Komoltsev:2021jzg}, and that all physical \eoss\ are causal, support $2 M_\odot$ NSs, and are consistent with the tidal-deformability from GW170817. We find that within our PT setup all three constraints -- chiral EFT, astrophysics, and pQCD -- provide unique constraints to the set of allowed \eoss.

Our derived region of possible PTs covers nearly all individual \eoss\ considered in previous works on PTs and their inclusion in merger simulations, but is significantly larger and includes several scenarios that have not yet been explored in merger simulations. Moreover, we find that first-order PTs can extend the radius ranges of NSs to larger radii, but these are only compatible with astrophysics constraints if the PT begins early. Note that these solutions are also broadly consistent with the radius constraints from NICER \citep{Miller:2021qha,Riley:2021pdl}. Our \eos\ ensemble also includes twin-star solutions, although their parameter space is very small, with the solutions just barely passing current astrophysical constraints. As expected, sizable cores, i.e., with a significant part of matter described by the second polytrope beyond the phase transition, mainly occur in stars with large masses.

Let us finally comment that the \eos\ ansatz used in this work is exploratory, so it is to be expected that the derived bounds for the \eoss\ and PT parameters can increase somewhat if one allows for more freedom in the \eos.
Nevertheless, our systematic study with explicit PTs points to important qualitative features for NS properties possible with strong PTs.

\section*{Acknowledgments}

We thank Yannick Dietz and Matthias Heinz for help with writing and optimizing the code to construct the NS sequences. We thank Sophia Han for sharing the data from \citet{Han:2018mtj}. This work is supported in part by the State of Hesse within the Research Cluster ELEMENTS (Project ID 500/10.006), the European Research Council (ERC) under the European Union's Horizon 2020 research and innovation programme (Grant Agreement Nos.~101020842 and 725369), and the Academy of Finland (Grant No.~1322507).

\bibliographystyle{aasjournal}
\bibliography{apJ_refs.bib}

\begin{thebibliography}{}
\expandafter\ifx\csname natexlab\endcsname\relax\def\natexlab#1{#1}\fi
\providecommand{\url}[1]{\href{#1}{#1}}
\providecommand{\dodoi}[1]{doi:~\href{http://doi.org/#1}{\nolinkurl{#1}}}
\providecommand{\doeprint}[1]{\href{http://ascl.net/#1}{\nolinkurl{http://ascl.net/#1}}}
\providecommand{\doarXiv}[1]{\href{https://arxiv.org/abs/#1}{\nolinkurl{https://arxiv.org/abs/#1}}}

\bibitem[{Abbott {et~al.}(2017)Abbott, Abbott, Abbott, Acernese, Ackley, Adams,
  Adams, Addesso, Adhikari, Adya, Affeldt, Afrough, Agarwal, Agathos, Agatsuma,
  Aggarwal, Aguiar, Aiello, Ain, Ajith, Allen, Allen, Allocca, Altin, Amato,
  Ananyeva, Anderson, Anderson, Angelova, Antier, Appert, Arai, Araya, Areeda,
  Arnaud, Arun, Ascenzi, Ashton, Ast, Aston, Astone, Atallah, Aufmuth, Aulbert,
  AultONeal, Austin, Avila-Alvarez, Babak, Bacon, Bader, Bae, Bailes, Baker,
  Baldaccini, Ballardin, Ballmer, Banagiri, Barayoga, Barclay, Barish, Barker,
  Barkett, Barone, Barr, Barsotti, Barsuglia, Barta, Barthelmy, Bartlett,
  Bartos, Bassiri, Basti, Batch, Bawaj, Bayley, Bazzan, B\'ecsy, Beer, Bejger,
  Belahcene, Bell, Berger, Bergmann, Bernuzzi, Bero, Berry, Bersanetti,
  Bertolini, Betzwieser, Bhagwat, Bhandare, Bilenko, Billingsley, Billman,
  Birch, Birney, Birnholtz, Biscans, Biscoveanu, Bisht, Bitossi, Biwer,
  Bizouard, Blackburn, Blackman, Blair, Blair, Blair, Bloemen, Bock, Bode,
  Boer, Bogaert, Bohe, Bondu, Bonilla, Bonnand, Boom, Bork, Boschi, Bose,
  Bossie, Bouffanais, Bozzi, Bradaschia, Brady, Branchesi, Brau, Briant,
  Brillet, Brinkmann, Brisson, Brockill, Broida, Brooks, Brown, Brown, Brunett,
  Buchanan, Buikema, Bulik, Bulten, Buonanno, Buskulic, Buy, Byer, Cabero,
  Cadonati, Cagnoli, Cahillane, Calder\'on~Bustillo, Callister, Calloni, Camp,
  Canepa, Canizares, Cannon, Cao, Cao, Capano, Capocasa, Carbognani, Caride,
  Carney, Carullo, Casanueva~Diaz, Casentini, Caudill, Cavagli\`a, Cavalier,
  Cavalieri, Cella, Cepeda, Cerd\'a-Dur\'an, Cerretani, Cesarini, Chamberlin,
  Chan, Chao, Charlton, Chase, Chassande-Mottin, Chatterjee, Chatziioannou,
  Cheeseboro, Chen, Chen, Chen, Cheng, Chia, Chincarini, Chiummo, Chmiel, Cho,
  Cho, Chow, Christensen, Chu, Chua, Chua, Chung, Chung, Ciani, Ciolfi,
  Cirelli, Cirone, Clara, Clark, Clearwater, Cleva, Cocchieri, Coccia, Cohadon,
  Cohen, Colla, Collette, Cominsky, Constancio, Conti, Cooper, Corban, Corbitt,
  Cordero-Carri\'on, Corley, Cornish, Corsi, Cortese, Costa, Coughlin,
  Coughlin, Coulon, Countryman, Couvares, Covas, Cowan, Coward, Cowart, Coyne,
  Coyne, Creighton, Creighton, Cripe, Crowder, Cullen, Cumming, Cunningham,
  Cuoco, Dal~Canton, D\'alya, Danilishin, D'Antonio, Danzmann, Dasgupta,
  Da~Silva~Costa, Dattilo, Dave, Davier, Davis, Daw, Day, De, DeBra, Degallaix,
  De~Laurentis, Del\'eglise, Del~Pozzo, Demos, Denker, Dent, De~Pietri,
  Dergachev, De~Rosa, DeRosa, De~Rossi, DeSalvo, de~Varona, Devenson,
  Dhurandhar, D\'{\i}az, Dietrich, Di~Fiore, Di~Giovanni, Di~Girolamo,
  Di~Lieto, Di~Pace, Di~Palma, Di~Renzo, Doctor, Dolique, Donovan, Dooley,
  Doravari, Dorrington, Douglas, Dovale~\'Alvarez, Downes, Drago,
  Dreissigacker, Driggers, Du, Ducrot, Dudi, Dupej, Dwyer, Edo, Edwards,
  Effler, Eggenstein, Ehrens, Eichholz, Eikenberry, Eisenstein, Essick,
  Estevez, Etienne, Etzel, Evans, Evans, Factourovich, Fafone, Fair, Fairhurst,
  Fan, Farinon, Farr, Farr, Fauchon-Jones, Favata, Fays, Fee, Fehrmann, Feicht,
  Fejer, Fernandez-Galiana, Ferrante, Ferreira, Ferrini, Fidecaro, Finstad,
  Fiori, Fiorucci, Fishbach, Fisher, Fitz-Axen, Flaminio, Fletcher, Fong, Font,
  Forsyth, Forsyth, Fournier, Frasca, Frasconi, Frei, Freise, Frey, Frey,
  Fries, Fritschel, Frolov, Fulda, Fyffe, Gabbard, Gadre, Gaebel, Gair,
  Gammaitoni, Ganija, Gaonkar, Garcia-Quiros, Garufi, Gateley, Gaudio, Gaur,
  Gayathri, Gehrels, Gemme, Genin, Gennai, George, George, Gergely, Germain,
  Ghonge, Ghosh, Ghosh, Ghosh, Giaime, Giardina, Giazotto, Gill, Glover, Goetz,
  Goetz, Gomes, Goncharov, Gonz\'alez, Gonzalez~Castro, Gopakumar, Gorodetsky,
  Gossan, Gosselin, Gouaty, Grado, Graef, Granata, Grant, Gras, Gray, Greco,
  Green, Gretarsson, Groot, Grote, Grunewald, Gruning, Guidi, Guo, Gupta,
  Gupta, Gushwa, Gustafson, Gustafson, Halim, Hall, Hall, Hamilton, Hammond,
  Haney, Hanke, Hanks, Hanna, Hannam, Hannuksela, Hanson, Hardwick, Harms,
  Harry, Harry, Hart, Haster, Haughian, Healy, Heidmann, Heintze, Heitmann,
  Hello, Hemming, Hendry, Heng, Hennig, Heptonstall, Heurs, Hild, Hinderer, Ho,
  Hoak, Hofman, Holt, Holz, Hopkins, Horst, Hough, Houston, Howell, Hreibi, Hu,
  Huerta, Huet, Hughey, Husa, Huttner, Huynh-Dinh, Indik, Inta, Intini, Isa,
  Isac, Isi, Iyer, Izumi, Jacqmin, Jani, Jaranowski, Jawahar,
  Jim\'enez-Forteza, Johnson, Johnson-McDaniel, Jones, Jones, Jonker, Ju,
  Junker, Kalaghatgi, Kalogera, Kamai, Kandhasamy, Kang, Kanner, Kapadia,
  Karki, Karvinen, Kasprzack, Kastaun, Katolik, Katsavounidis, Katzman, Kaufer,
  Kawabe, K\'ef\'elian, Keitel, Kemball, Kennedy, Kent, Key, Khalili, Khan,
  Khan, Khan, Khazanov, Kijbunchoo, Kim, Kim, Kim, Kim, Kim, Kim, Kimbrell,
  King, King, Kinley-Hanlon, Kirchhoff, Kissel, Kleybolte, Klimenko, Knowles,
  Koch, Koehlenbeck, Koley, Kondrashov, Kontos, Korobko, Korth, Kowalska,
  Kozak, Kr\"amer, Kringel, Krishnan, Kr\'olak, Kuehn, Kumar, Kumar, Kumar,
  Kuo, Kutynia, Kwang, Lackey, Lai, Landry, Lang, Lange, Lantz, Lanza, Larson,
  Lartaux-Vollard, Lasky, Laxen, Lazzarini, Lazzaro, Leaci, Leavey, Lee, Lee,
  Lee, Lee, Lee, Lehmann, Lenon, Leon, Leonardi, Leroy, Letendre, Levin, Li,
  Linker, Littenberg, Liu, Liu, Lo, Lockerbie, London, Lord, Lorenzini,
  Loriette, Lormand, Losurdo, Lough, Lousto, Lovelace, L\"uck, Lumaca,
  Lundgren, Lynch, Ma, Macas, Macfoy, Machenschalk, MacInnis, Macleod, Maga\~na
  Hernandez, Maga\~na Sandoval, Maga\~na Zertuche, Magee, Majorana, Maksimovic,
  Man, Mandic, Mangano, Mansell, Manske, Mantovani, Marchesoni, Marion,
  M\'arka, M\'arka, Markakis, Markosyan, Markowitz, Maros, Marquina, Marsh,
  Martelli, Martellini, Martin, Martin, Martynov, Marx, Mason, Massera,
  Masserot, Massinger, Masso-Reid, Mastrogiovanni, Matas, Matichard, Matone,
  Mavalvala, Mazumder, McCarthy, McClelland, McCormick, McCuller, McGuire,
  McIntyre, McIver, McManus, McNeill, McRae, McWilliams, Meacher, Meadors,
  Mehmet, Meidam, Mejuto-Villa, Melatos, Mendell, Mercer, Merilh, Merzougui,
  Meshkov, Messenger, Messick, Metzdorff, Meyers, Miao, Michel, Middleton,
  Mikhailov, Milano, Miller, Miller, Miller, Millhouse, Milovich-Goff,
  Minazzoli, Minenkov, Ming, Mishra, Mitra, Mitrofanov, Mitselmakher,
  Mittleman, Moffa, Moggi, Mogushi, Mohan, Mohapatra, Molina, Montani, Moore,
  Moraru, Moreno, Morisaki, Morriss, Mours, Mow-Lowry, Mueller, Muir,
  Mukherjee, Mukherjee, Mukherjee, Mukund, Mullavey, Munch, Mu\~niz, Muratore,
  Murray, Nagar, Napier, Nardecchia, Naticchioni, Nayak, Neilson, Nelemans,
  Nelson, Nery, Neunzert, Nevin, Newport, Newton, Ng, Nguyen, Nguyen, Nichols,
  Nielsen, Nissanke, Nitz, Noack, Nocera, Nolting, North, Nuttall, Oberling,
  O'Dea, Ogin, Oh, Oh, Ohme, Okada, Oliver, Oppermann, Oram, O'Reilly,
  Ormiston, Ortega, O'Shaughnessy, Ossokine, Ottaway, Overmier, Owen, Pace,
  Page, Page, Pai, Pai, Palamos, Palashov, Palomba, Pal-Singh, Pan, Pan, Pang,
  Pang, Pankow, Pannarale, Pant, Paoletti, Paoli, Papa, Parida, Parker,
  Pascucci, Pasqualetti, Passaquieti, Passuello, Patil, Patricelli, Pearlstone,
  Pedraza, Pedurand, Pekowsky, Pele, Penn, Perez, Perreca, Perri, Pfeiffer,
  Phelps, Piccinni, Pichot, Piergiovanni, Pierro, Pillant, Pinard, Pinto,
  Pirello, Pitkin, Poe, Poggiani, Popolizio, Porter, Post, Powell, Prasad,
  Pratt, Pratten, Predoi, Prestegard, Prijatelj, Principe, Privitera, Prix,
  Prodi, Prokhorov, Puncken, Punturo, Puppo, P\"urrer, Qi, Quetschke, Quintero,
  Quitzow-James, Raab, Rabeling, Radkins, Raffai, Raja, Rajan, Rajbhandari,
  Rakhmanov, Ramirez, Ramos-Buades, Rapagnani, Raymond, Razzano, Read,
  Regimbau, Rei, Reid, Reitze, Ren, Reyes, Ricci, Ricker, Rieger, Riles, Rizzo,
  Robertson, Robie, Robinet, Rocchi, Rolland, Rollins, Roma, Romano, Romano,
  Romel, Romie, Rosi\ifmmode~\acute{n}\else \'{n}\fi{}ska, Ross, Rowan,
  R\"udiger, Ruggi, Rutins, Ryan, Sachdev, Sadecki, Sadeghian, Sakellariadou,
  Salconi, Saleem, Salemi, Samajdar, Sammut, Sampson, Sanchez, Sanchez,
  Sanchis-Gual, Sandberg, Sanders, Sassolas, Sathyaprakash, Saulson, Sauter,
  Savage, Sawadsky, Schale, Scheel, Scheuer, Schmidt, Schmidt, Schnabel,
  Schofield, Sch\"onbeck, Schreiber, Schuette, Schulte, Schutz, Schwalbe,
  Scott, Scott, Seidel, Sellers, Sengupta, Sentenac, Sequino, Sergeev,
  Shaddock, Shaffer, Shah, Shahriar, Shaner, Shao, Shapiro, Shawhan, Sheperd,
  Shoemaker, Shoemaker, Siellez, Siemens, Sieniawska, Sigg, Silva, Singer,
  Singh, Singhal, Sintes, Slagmolen, Smith, Smith, Smith, Somala, Son,
  Sonnenberg, Sorazu, Sorrentino, Souradeep, Spencer, Srivastava, Staats,
  Staley, Steinke, Steinlechner, Steinlechner, Steinmeyer, Stevenson, Stone,
  Stops, Strain, Stratta, Strigin, Strunk, Sturani, Stuver, Summerscales, Sun,
  Sunil, Suresh, Sutton, Swinkels, Szczepa\ifmmode~\acute{n}\else
  \'{n}\fi{}czyk, Tacca, Tait, Talbot, Talukder, Tanner, T\'apai, Taracchini,
  Tasson, Taylor, Taylor, Tewari, Theeg, Thies, Thomas, Thomas, Thomas, Thorne,
  Thorne, Thrane, Tiwari, Tiwari, Tokmakov, Toland, Tonelli, Tornasi,
  Torres-Forn\'e, Torrie, T\"oyr\"a, Travasso, Traylor, Trinastic, Tringali,
  Trozzo, Tsang, Tse, Tso, Tsukada, Tsuna, Tuyenbayev, Ueno, Ugolini,
  Unnikrishnan, Urban, Usman, Vahlbruch, Vajente, Valdes, Vallisneri, van
  Bakel, van Beuzekom, van~den Brand, Van Den~Broeck, Vander-Hyde, van~der
  Schaaf, van Heijningen, van Veggel, Vardaro, Varma, Vass, Vas\'uth, Vecchio,
  Vedovato, Veitch, Veitch, Venkateswara, Venugopalan, Verkindt, Vetrano,
  Vicer\'e, Viets, Vinciguerra, Vine, Vinet, Vitale, Vo, Vocca, Vorvick,
  Vyatchanin, Wade, Wade, Wade, Walet, Walker, Wallace, Walsh, Wang, Wang,
  Wang, Wang, Wang, Ward, Warner, Was, Watchi, Weaver, Wei, Weinert, Weinstein,
  Weiss, Wen, Wessel, We\ss{}els, Westerweck, Westphal, Wette, Whelan,
  Whitcomb, Whiting, Whittle, Wilken, Williams, Williams, Williamson, Willis,
  Willke, Wimmer, Winkler, Wipf, Wittel, Woan, Woehler, Wofford, Wong, Worden,
  Wright, Wu, Wysocki, Xiao, Yamamoto, Yancey, Yang, Yap, Yazback, Yu, Yu,
  Yvert, Zadro\ifmmode~\dot{z}\else \.{z}\fi{}ny, Zanolin, Zelenova, Zendri,
  Zevin, Zhang, Zhang, Zhang, Zhang, Zhao, Zhou, Zhou, Zhu, Zhu, Zimmerman,
  Zucker, \& Zweizig}]{LIGOScientific:2017vwq}
Abbott, B.~P., Abbott, R., Abbott, T.~D., {et~al.} 2017, Phys. Rev. Lett., 119,
  161101, \dodoi{10.1103/PhysRevLett.119.161101}

\bibitem[{Abbott {et~al.}(2018)Abbott, Abbott, Abbott, Acernese, Ackley, Adams,
  Adams, Addesso, Adhikari, Adya, Affeldt, Agarwal, Agathos, Agatsuma,
  Aggarwal, Aguiar, Aiello, Ain, Ajith, Allen, Allen, Allocca, Aloy, Altin,
  Amato, Ananyeva, Anderson, Anderson, Angelova, Antier, Appert, Arai, Araya,
  Areeda, Ar\`ene, Arnaud, Arun, Ascenzi, Ashton, Ast, Aston, Astone, Atallah,
  Aubin, Aufmuth, Aulbert, AultONeal, Austin, Avila-Alvarez, Babak, Bacon,
  Badaracco, Bader, Bae, Baker, Baldaccini, Ballardin, Ballmer, Banagiri,
  Barayoga, Barclay, Barish, Barker, Barkett, Barnum, Barone, Barr, Barsotti,
  Barsuglia, Barta, Bartlett, Bartos, Bassiri, Basti, Batch, Bawaj, Bayley,
  Bazzan, B\'ecsy, Beer, Bejger, Belahcene, Bell, Beniwal, Bensch, Berger,
  Bergmann, Bernuzzi, Bero, Berry, Bersanetti, Bertolini, Betzwieser, Bhandare,
  Bilenko, Bilgili, Billingsley, Billman, Birch, Birney, Birnholtz, Biscans,
  Biscoveanu, Bisht, Bitossi, Bizouard, Blackburn, Blackman, Blair, Blair,
  Blair, Bloemen, Bock, Bode, Boer, Boetzel, Bogaert, Bohe, Bondu, Bonilla,
  Bonnand, Booker, Boom, Booth, Bork, Boschi, Bose, Bossie, Bossilkov, Bosveld,
  Bouffanais, Bozzi, Bradaschia, Brady, Bramley, Branchesi, Brau, Briant,
  Brighenti, Brillet, Brinkmann, Brisson, Brockill, Brooks, Brown, Brunett,
  Buchanan, Buikema, Bulik, Bulten, Buonanno, Buskulic, Buy, Byer, Cabero,
  Cadonati, Cagnoli, Cahillane, Calder\'on~Bustillo, Callister, Calloni, Camp,
  Canepa, Canizares, Cannon, Cao, Cao, Capano, Capocasa, Carbognani, Caride,
  Carney, Carullo, Casanueva~Diaz, Casentini, Caudill, Cavagli\`a, Cavalier,
  Cavalieri, Cella, Cepeda, Cerd\'a-Dur\'an, Cerretani, Cesarini, Chaibi,
  Chamberlin, Chan, Chao, Charlton, Chase, Chassande-Mottin, Chatterjee,
  Chatziioannou, Cheeseboro, Chen, Chen, Chen, Cheng, Chia, Chincarini,
  Chiummo, Chmiel, Cho, Cho, Chow, Christensen, Chu, Chua, Chua, Chung, Chung,
  Ciani, Ciobanu, Ciolfi, Cipriano, Cirelli, Cirone, Clara, Clark, Clearwater,
  Cleva, Cocchieri, Coccia, Cohadon, Cohen, Colla, Collette, Collins, Cominsky,
  Constancio, Conti, Cooper, Corban, Corbitt, Cordero-Carri\'on, Corley,
  Cornish, Corsi, Cortese, Costa, Cotesta, Coughlin, Coughlin, Coulon,
  Countryman, Couvares, Covas, Cowan, Coward, Cowart, Coyne, Coyne, Creighton,
  Creighton, Cripe, Crowder, Cullen, Cumming, Cunningham, Cuoco, Canton,
  D\'alya, Danilishin, D'Antonio, Danzmann, Dasgupta, Da~Silva~Costa, Dattilo,
  Dave, Davier, Davis, Daw, Day, DeBra, Deenadayalan, Degallaix, De~Laurentis,
  Del\'eglise, Del~Pozzo, Demos, Denker, Dent, De~Pietri, Derby, Dergachev,
  De~Rosa, De~Rossi, DeSalvo, de~Varona, Dhurandhar, D\'{\i}az, Dietrich,
  Di~Fiore, Di~Giovanni, Di~Girolamo, Di~Lieto, Ding, Di~Pace, Di~Palma,
  Di~Renzo, Dmitriev, Doctor, Dolique, Donovan, Dooley, Doravari, Dorrington,
  Dovale~\'Alvarez, Downes, Drago, Dreissigacker, Driggers, Du, Dupej, Dwyer,
  Easter, Edo, Edwards, Effler, Eggenstein, Ehrens, Eichholz, Eikenberry,
  Eisenmann, Eisenstein, Essick, Estelles, Estevez, Etienne, Etzel, Evans,
  Evans, Fafone, Fair, Fairhurst, Fan, Farinon, Farr, Farr, Fauchon-Jones,
  Favata, Fays, Fee, Fehrmann, Feicht, Fejer, Feng, Fernandez-Galiana,
  Ferrante, Ferreira, Ferrini, Fidecaro, Fiori, Fiorucci, Fishbach, Fisher,
  Fishner, Fitz-Axen, Flaminio, Fletcher, Fong, Font, Forsyth, Forsyth,
  Fournier, Frasca, Frasconi, Frei, Freise, Frey, Frey, Fritschel, Frolov,
  Fulda, Fyffe, Gabbard, Gadre, Gaebel, Gair, Gammaitoni, Ganija, Gaonkar,
  Garcia, Garc\'{\i}a-Quir\'os, Garufi, Gateley, Gaudio, Gaur, Gayathri, Gemme,
  Genin, Gennai, George, George, Gergely, Germain, Ghonge, Ghosh, Ghosh, Ghosh,
  Giacomazzo, Giaime, Giardina, Giazotto, Gill, Giordano, Glover, Goetz, Goetz,
  Goncharov, Gonz\'alez, Gonzalez~Castro, Gopakumar, Gorodetsky, Gossan,
  Gosselin, Gouaty, Grado, Graef, Granata, Grant, Gras, Gray, Greco, Green,
  Green, Gretarsson, Groot, Grote, Grunewald, Gruning, Guidi, Gulati, Guo,
  Gupta, Gupta, Gushwa, Gustafson, Gustafson, Halim, Hall, Hall, Hamilton,
  Hamilton, Hammond, Haney, Hanke, Hanks, Hanna, Hannam, Hannuksela, Hanson,
  Hardwick, Harms, Harry, Harry, Hart, Haster, Haughian, Healy, Heidmann,
  Heintze, Heitmann, Hello, Hemming, Hendry, Heng, Hennig, Heptonstall,
  Hernandez, Heurs, Hild, Hinderer, Ho, Hoak, Hochheim, Hofman, Holland, Holt,
  Holz, Hopkins, Horst, Hough, Houston, Howell, Hreibi, Huerta, Huet, Hughey,
  Hulko, Husa, Huttner, Huynh-Dinh, Iess, Indik, Ingram, Inta, Intini, Irwin,
  Isa, Isac, Isi, Iyer, Izumi, Jacqmin, Jani, Jaranowski, Johnson, Johnson,
  Jones, Jones, Jonker, Ju, Junker, Kalaghatgi, Kalogera, Kamai, Kandhasamy,
  Kang, Kanner, Kapadia, Karki, Karvinen, Kasprzack, Katolik, Katsanevas,
  Katsavounidis, Katzman, Kaufer, Kawabe, Keerthana, K\'ef\'elian, Keitel,
  Kemball, Kennedy, Key, Khalili, Khamesra, Khan, Khan, Khan, Khan, Khazanov,
  Kijbunchoo, Kim, Kim, Kim, Kim, Kim, Kim, King, King, Kinley-Hanlon,
  Kirchhoff, Kissel, Kleybolte, Klimenko, Knowles, Koch, Koehlenbeck, Koley,
  Kondrashov, Kontos, Korobko, Korth, Kowalska, Kozak, Kr\"amer, Kringel,
  Krishnan, Kr\'olak, Kuehn, Kumar, Kumar, Kumar, Kuo, Kutynia, Kwang, Lackey,
  Lai, Landry, Landry, Lang, Lange, Lantz, Lanza, Lartaux-Vollard, Lasky,
  Laxen, Lazzarini, Lazzaro, Leaci, Leavey, Lee, Lee, Lee, Lee, Lee, Lehmann,
  Lenon, Leonardi, Leroy, Letendre, Levin, Li, Li, Li, Linker, Littenberg, Liu,
  Liu, Lo, Lockerbie, London, Longo, Lorenzini, Loriette, Lormand, Losurdo,
  Lough, Lousto, Lovelace, L\"uck, Lumaca, Lundgren, Lynch, Ma, Macas, Macfoy,
  Machenschalk, MacInnis, Macleod, Maga\~na Hernandez, Maga\~na Sandoval,
  Maga\~na Zertuche, Magee, Majorana, Maksimovic, Man, Mandic, Mangano,
  Mansell, Manske, Mantovani, Marchesoni, Marion, M\'arka, M\'arka, Markakis,
  Markosyan, Markowitz, Maros, Marquina, Martelli, Martellini, Martin, Martin,
  Martynov, Mason, Massera, Masserot, Massinger, Masso-Reid, Mastrogiovanni,
  Matas, Matichard, Matone, Mavalvala, Mazumder, McCann, McCarthy, McClelland,
  McCormick, McCuller, McGuire, McIver, McManus, McRae, McWilliams, Meacher,
  Meadors, Mehmet, Meidam, Mejuto-Villa, Melatos, Mendell, Mendoza-Gandara,
  Mercer, Mereni, Merilh, Merzougui, Meshkov, Messenger, Messick, Metzdorff,
  Meyers, Miao, Michel, Middleton, Mikhailov, Milano, Miller, Miller, Miller,
  Miller, Millhouse, Mills, Milovich-Goff, Minazzoli, Minenkov, Ming, Mishra,
  Mitra, Mitrofanov, Mitselmakher, Mittleman, Moffa, Mogushi, Mohan, Mohapatra,
  Montani, Moore, Moraru, Moreno, Morisaki, Mours, Mow-Lowry, Mueller, Muir,
  Mukherjee, Mukherjee, Mukherjee, Mukund, Mullavey, Munch, Mu\~niz, Muratore,
  Murray, Nagar, Napier, Nardecchia, Naticchioni, Nayak, Neilson, Nelemans,
  Nelson, Nery, Neunzert, Nevin, Newport, Ng, Ng, Nguyen, Nguyen, Nichols,
  Nielsen, Nissanke, Nitz, Nocera, Nolting, North, Nuttall, Obergaulinger,
  Oberling, O'Brien, O'Dea, Ogin, Oh, Oh, Ohme, Ohta, Okada, Oliver, Oppermann,
  Oram, O'Reilly, Ormiston, Ortega, O'Shaughnessy, Ossokine, Ottaway, Overmier,
  Owen, Pace, Pagano, Page, Page, Pai, Pai, Palamos, Palashov, Palomba,
  Pal-Singh, Pan, Pan, Pang, Pang, Pankow, Pannarale, Pant, Paoletti, Paoli,
  Papa, Parida, Parker, Pascucci, Pasqualetti, Passaquieti, Passuello, Patil,
  Patricelli, Pearlstone, Pedersen, Pedraza, Pedurand, Pekowsky, Pele, Penn,
  Perego, Perez, Perreca, Perri, Pfeiffer, Phelps, Phukon, Piccinni, Pichot,
  Piergiovanni, Pierro, Pillant, Pinard, Pinto, Pirello, Pitkin, Poggiani,
  Popolizio, Porter, Possenti, Post, Powell, Prasad, Pratt, Pratten, Predoi,
  Prestegard, Principe, Privitera, Prodi, Prokhorov, Puncken, Punturo, Puppo,
  P\"urrer, Qi, Quetschke, Quintero, Quitzow-James, Raab, Rabeling, Radkins,
  Raffai, Raja, Rajan, Rajbhandari, Rakhmanov, Ramirez, Ramos-Buades, Rana,
  Rapagnani, Raymond, Razzano, Read, Regimbau, Rei, Reid, Reitze, Ren, Ricci,
  Ricker, Riemenschneider, Riles, Rizzo, Robertson, Robie, Robinet, Robson,
  Rocchi, Rolland, Rollins, Roma, Romano, Romel, Romie,
  Rosi\ifmmode~\acute{n}\else \'{n}\fi{}ska, Ross, Rowan, R\"udiger, Ruggi,
  Rutins, Ryan, Sachdev, Sadecki, Sakellariadou, Salconi, Saleem, Salemi,
  Samajdar, Sammut, Sampson, Sanchez, Sanchez, Sanchis-Gual, Sandberg, Sanders,
  Sarin, Sassolas, Sathyaprakash, Saulson, Sauter, Savage, Sawadsky, Schale,
  Scheel, Scheuer, Schmidt, Schnabel, Schofield, Sch\"onbeck, Schreiber,
  Schuette, Schulte, Schutz, Schwalbe, Scott, Scott, Seidel, Sellers, Sengupta,
  Sentenac, Sequino, Sergeev, Setyawati, Shaddock, Shaffer, Shah, Shahriar,
  Shaner, Shao, Shapiro, Shawhan, Shen, Shoemaker, Shoemaker, Siellez, Siemens,
  Sieniawska, Sigg, Silva, Singer, Singh, Singhal, Sintes, Slagmolen,
  Slaven-Blair, Smith, Smith, Smith, Somala, Son, Sorazu, Sorrentino,
  Souradeep, Spencer, Srivastava, Staats, Steinke, Steinlechner, Steinlechner,
  Steinmeyer, Steltner, Stevenson, Stocks, Stone, Stops, Strain, Stratta,
  Strigin, Strunk, Sturani, Stuver, Summerscales, Sun, Sunil, Suresh, Sutton,
  Swinkels, Szczepa\ifmmode~\acute{n}\else \'{n}\fi{}czyk, Tacca, Tait, Talbot,
  Talukder, Tanner, T\'apai, Taracchini, Tasson, Taylor, Taylor, Tewari, Theeg,
  Thies, Thomas, Thomas, Thomas, Thorne, Thrane, Tiwari, Tiwari, Tokmakov,
  Toland, Tonelli, Tornasi, Torres-Forn\'e, Torrie, T\"oyr\"a, Travasso,
  Traylor, Trinastic, Tringali, Trovato, Trozzo, Tsang, Tse, Tso, Tsuna,
  Tsukada, Tuyenbayev, Ueno, Ugolini, Urban, Usman, Vahlbruch, Vajente, Valdes,
  van Bakel, van Beuzekom, van~den Brand, Van Den~Broeck, Vander-Hyde, van~der
  Schaaf, van Heijningen, van Veggel, Vardaro, Varma, Vass, Vas\'uth, Vecchio,
  Vedovato, Veitch, Veitch, Venkateswara, Venugopalan, Verkindt, Vetrano,
  Vicer\'e, Viets, Vinciguerra, Vine, Vinet, Vitale, Vo, Vocca, Vorvick,
  Vyatchanin, Wade, Wade, Wade, Walet, Walker, Wallace, Walsh, Wang, Wang,
  Wang, Wang, Wang, Ward, Warner, Was, Watchi, Weaver, Wei, Weinert, Weinstein,
  Weiss, Wellmann, Wen, Wessel, We\ss{}els, Westerweck, Wette, Whelan, Whiting,
  Whittle, Wilken, Williams, Williams, Williamson, Willis, Willke, Wimmer,
  Winkler, Wipf, Wittel, Woan, Woehler, Wofford, Wong, Worden, Wright, Wu,
  Wysocki, Xiao, Yam, Yamamoto, Yancey, Yang, Yap, Yazback, Yu, Yu, Yvert,
  Zadro\ifmmode~\dot{z}\else \.{z}\fi{}ny, Zanolin, Zelenova, Zendri, Zevin,
  Zhang, Zhang, Zhang, Zhang, Zhang, Zhao, Zhou, Zhou, Zhu, Zhu, Zimmerman,
  Zlochower, Zucker, \& Zweizig}]{LIGOScientific:2018cki}
---. 2018, Phys. Rev. Lett., 121, 161101,
  \dodoi{10.1103/PhysRevLett.121.161101}

\bibitem[{Abbott {et~al.}(2019)Abbott, Abbott, Abbott, Acernese, Ackley, Adams,
  Adams, Addesso, Adhikari, Adya, Affeldt, Agarwal, Agathos, Agatsuma,
  Aggarwal, Aguiar, Aiello, Ain, Ajith, Allen, Allen, Allocca, Aloy, Altin,
  Amato, Ananyeva, Anderson, Anderson, Angelova, Antier, Appert, Arai, Araya,
  Areeda, Ar\`ene, Arnaud, Arun, Ascenzi, Ashton, Ast, Aston, Astone, Atallah,
  Aubin, Aufmuth, Aulbert, AultONeal, Austin, Avila-Alvarez, Babak, Bacon,
  Badaracco, Bader, Bae, Baker, Baldaccini, Ballardin, Ballmer, Banagiri,
  Barayoga, Barclay, Barish, Barker, Barkett, Barnum, Barone, Barr, Barsotti,
  Barsuglia, Barta, Bartlett, Bartos, Bassiri, Basti, Batch, Bawaj, Bayley,
  Bazzan, B\'ecsy, Beer, Bejger, Belahcene, Bell, Beniwal, Bensch, Berger,
  Bergmann, Bernuzzi, Bero, Berry, Bersanetti, Bertolini, Betzwieser, Bhandare,
  Bilenko, Bilgili, Billingsley, Billman, Birch, Birney, Birnholtz, Biscans,
  Biscoveanu, Bisht, Bitossi, Bizouard, Blackburn, Blackman, Blair, Blair,
  Blair, Bloemen, Bock, Bode, Boer, Boetzel, Bogaert, Bohe, Bondu, Bonilla,
  Bonnand, Booker, Boom, Booth, Bork, Boschi, Bose, Bossie, Bossilkov, Bosveld,
  Bouffanais, Bozzi, Bradaschia, Brady, Bramley, Branchesi, Brau, Briant,
  Brighenti, Brillet, Brinkmann, Brisson, Brockill, Brooks, Brown, Brunett,
  Buchanan, Buikema, Bulik, Bulten, Buonanno, Buskulic, Buy, Byer, Cabero,
  Cadonati, Cagnoli, Cahillane, Bustillo, Callister, Calloni, Camp, Canepa,
  Canizares, Cannon, Cao, Cao, Capano, Capocasa, Carbognani, Caride, Carney,
  Carullo, Diaz, Casentini, Caudill, Cavagli\`a, Cavalier, Cavalieri, Cella,
  Cepeda, Cerd\'a-Dur\'an, Cerretani, Cesarini, Chaibi, Chamberlin, Chan, Chao,
  Charlton, Chase, Chassande-Mottin, Chatterjee, Chatziioannou, Cheeseboro,
  Chen, Chen, Chen, Cheng, Chia, Chincarini, Chiummo, Chmiel, Cho, Cho, Chow,
  Christensen, Chu, Chua, Chua, Chung, Chung, Ciani, Ciobanu, Ciolfi, Cipriano,
  Cirelli, Cirone, Clara, Clark, Clearwater, Cleva, Cocchieri, Coccia, Cohadon,
  Cohen, Colla, Collette, Collins, Cominsky, Constancio, Conti, Cooper, Corban,
  Corbitt, Cordero-Carri\'on, Corley, Cornish, Corsi, Cortese, Costa, Cotesta,
  Coughlin, Coughlin, Coulon, Countryman, Couvares, Covas, Cowan, Coward,
  Cowart, Coyne, Coyne, Creighton, Creighton, Cripe, Crowder, Cullen, Cumming,
  Cunningham, Cuoco, Canton, D\'alya, Danilishin, D'Antonio, Danzmann,
  Dasgupta, Costa, Dattilo, Dave, Davier, Davis, Daw, Day, DeBra, Deenadayalan,
  Degallaix, De~Laurentis, Del\'eglise, Del~Pozzo, Demos, Denker, Dent,
  De~Pietri, Derby, Dergachev, De~Rosa, De~Rossi, DeSalvo, de~Varona,
  Dhurandhar, D\'{\i}az, Dietrich, Di~Fiore, Di~Giovanni, Di~Girolamo,
  Di~Lieto, Ding, Di~Pace, Di~Palma, Di~Renzo, Dmitriev, Doctor, Dolique,
  Donovan, Dooley, Doravari, Dorrington, \'Alvarez, Downes, Drago,
  Dreissigacker, Driggers, Du, Dudi, Dupej, Dwyer, Easter, Edo, Edwards,
  Effler, Eggenstein, Ehrens, Eichholz, Eikenberry, Eisenmann, Eisenstein,
  Essick, Estelles, Estevez, Etienne, Etzel, Evans, Evans, Fafone, Fair,
  Fairhurst, Fan, Farinon, Farr, Farr, Fauchon-Jones, Favata, Fays, Fee,
  Fehrmann, Feicht, Fejer, Feng, Fernandez-Galiana, Ferrante, Ferreira,
  Ferrini, Fidecaro, Fiori, Fiorucci, Fishbach, Fisher, Fishner, Fitz-Axen,
  Flaminio, Fletcher, Fong, Font, Forsyth, Forsyth, Fournier, Frasca, Frasconi,
  Frei, Freise, Frey, Frey, Fritschel, Frolov, Fulda, Fyffe, Gabbard, Gadre,
  Gaebel, Gair, Gammaitoni, Ganija, Gaonkar, Garcia, Garc\'{\i}a-Quir\'os,
  Garufi, Gateley, Gaudio, Gaur, Gayathri, Gemme, Genin, Gennai, George,
  George, Gergely, Germain, Ghonge, Ghosh, Ghosh, Ghosh, Giacomazzo, Giaime,
  Giardina, Giazotto, Gill, Giordano, Glover, Goetz, Goetz, Goncharov,
  Gonz\'alez, Castro, Gopakumar, Gorodetsky, Gossan, Gosselin, Gouaty, Grado,
  Graef, Granata, Grant, Gras, Gray, Greco, Green, Green, Gretarsson, Groot,
  Grote, Grunewald, Gruning, Guidi, Gulati, Guo, Gupta, Gupta, Gushwa,
  Gustafson, Gustafson, Halim, Hall, Hall, Hamilton, Hamilton, Hammond, Haney,
  Hanke, Hanks, Hanna, Hannam, Hannuksela, Hanson, Hardwick, Harms, Harry,
  Harry, Hart, Haster, Haughian, Healy, Heidmann, Heintze, Heitmann, Hello,
  Hemming, Hendry, Heng, Hennig, Heptonstall, Hernandez, Heurs, Hild, Hinderer,
  Hoak, Hochheim, Hofman, Holland, Holt, Holz, Hopkins, Horst, Hough, Houston,
  Howell, Hreibi, Huerta, Huet, Hughey, Hulko, Husa, Huttner, Huynh-Dinh, Iess,
  Indik, Ingram, Inta, Intini, Isa, Isac, Isi, Iyer, Izumi, Jacqmin, Jani,
  Jaranowski, Johnson, Johnson, Jones, Jones, Jonker, Ju, Junker, Kalaghatgi,
  Kalogera, Kamai, Kandhasamy, Kang, Kanner, Kapadia, Karki, Karvinen,
  Kasprzack, Kastaun, Katolik, Katsanevas, Katsavounidis, Katzman, Kaufer,
  Kawabe, Keerthana, K\'ef\'elian, Keitel, Kemball, Kennedy, Key, Khalili,
  Khamesra, Khan, Khan, Khan, Khan, Khazanov, Kijbunchoo, Kim, Kim, Kim, Kim,
  Kim, Kim, King, King, Kinley-Hanlon, Kirchhoff, Kissel, Kleybolte, Klimenko,
  Knowles, Koch, Koehlenbeck, Koley, Kondrashov, Kontos, Korobko, Korth,
  Kowalska, Kozak, Kr\"amer, Kringel, Krishnan, Kr\'olak, Kuehn, Kumar, Kumar,
  Kumar, Kuo, Kutynia, Kwang, Lackey, Lai, Landry, Landry, Lang, Lange, Lantz,
  Lanza, Lartaux-Vollard, Lasky, Laxen, Lazzarini, Lazzaro, Leaci, Leavey, Lee,
  Lee, Lee, Lee, Lee, Lehmann, Lenon, Leonardi, Leroy, Letendre, Levin, Li, Li,
  Li, Linker, Littenberg, Liu, Liu, Lo, Lockerbie, London, Longo, Lorenzini,
  Loriette, Lormand, Losurdo, Lough, Lousto, Lovelace, L\"uck, Lumaca,
  Lundgren, Lynch, Ma, Macas, Macfoy, Machenschalk, MacInnis, Macleod,
  Hernandez, Maga\~na Sandoval, Zertuche, Magee, Majorana, Maksimovic, Man,
  Mandic, Mangano, Mansell, Manske, Mantovani, Marchesoni, Marion, M\'arka,
  M\'arka, Markakis, Markosyan, Markowitz, Maros, Marquina, Martelli,
  Martellini, Martin, Martin, Martynov, Mason, Massera, Masserot, Massinger,
  Masso-Reid, Mastrogiovanni, Matas, Matichard, Matone, Mavalvala, Mazumder,
  McCann, McCarthy, McClelland, McCormick, McCuller, McGuire, McIver, McManus,
  McRae, McWilliams, Meacher, Meadors, Mehmet, Meidam, Mejuto-Villa, Melatos,
  Mendell, Mendoza-Gandara, Mercer, Mereni, Merilh, Merzougui, Meshkov,
  Messenger, Messick, Metzdorff, Meyers, Miao, Michel, Middleton, Mikhailov,
  Milano, Miller, Miller, Miller, Miller, Millhouse, Mills, Milovich-Goff,
  Minazzoli, Minenkov, Ming, Mishra, Mitra, Mitrofanov, Mitselmakher,
  Mittleman, Moffa, Mogushi, Mohan, Mohapatra, Montani, Moore, Moraru, Moreno,
  Morisaki, Mours, Mow-Lowry, Mueller, Muir, Mukherjee, Mukherjee, Mukherjee,
  Mukund, Mullavey, Munch, Mu\~niz, Muratore, Murray, Nagar, Napier,
  Nardecchia, Naticchioni, Nayak, Neilson, Nelemans, Nelson, Nery, Neunzert,
  Nevin, Newport, Ng, Ng, Nguyen, Nguyen, Nichols, Nielsen, Nissanke, Nitz,
  Nocera, Nolting, North, Nuttall, Obergaulinger, Oberling, O'Brien, O'Dea,
  Ogin, Oh, Oh, Ohme, Ohta, Okada, Oliver, Oppermann, Oram, O'Reilly, Ormiston,
  Ortega, O'Shaughnessy, Ossokine, Ottaway, Overmier, Owen, Pace, Pagano, Page,
  Page, Pai, Pai, Palamos, Palashov, Palomba, Pal-Singh, Pan, Pan, Pang, Pang,
  Pankow, Pannarale, Pant, Paoletti, Paoli, Papa, Parida, Parker, Pascucci,
  Pasqualetti, Passaquieti, Passuello, Patil, Patricelli, Pearlstone, Pedersen,
  Pedraza, Pedurand, Pekowsky, Pele, Penn, Perez, Perreca, Perri, Pfeiffer,
  Phelps, Phukon, Piccinni, Pichot, Piergiovanni, Pierro, Pillant, Pinard,
  Pinto, Pirello, Pitkin, Poggiani, Popolizio, Porter, Possenti, Post, Powell,
  Prasad, Pratt, Pratten, Predoi, Prestegard, Principe, Privitera, Prodi,
  Prokhorov, Puncken, Punturo, Puppo, P\"urrer, Qi, Quetschke, Quintero,
  Quitzow-James, Raab, Rabeling, Radkins, Raffai, Raja, Rajan, Rajbhandari,
  Rakhmanov, Ramirez, Ramos-Buades, Rana, Rapagnani, Raymond, Razzano, Read,
  Regimbau, Rei, Reid, Reitze, Ren, Ricci, Ricker, Riemenschneider, Riles,
  Rizzo, Robertson, Robie, Robinet, Robson, Rocchi, Rolland, Rollins, Roma,
  Romano, Romel, Romie, Rosi\ifmmode~\acute{n}\else \'{n}\fi{}ska, Ross, Rowan,
  R\"udiger, Ruggi, Rutins, Ryan, Sachdev, Sadecki, Sakellariadou, Salconi,
  Saleem, Salemi, Samajdar, Sammut, Sampson, Sanchez, Sanchez, Sanchis-Gual,
  Sandberg, Sanders, Sarin, Sassolas, Sathyaprakash, Saulson, Sauter, Savage,
  Sawadsky, Schale, Scheel, Scheuer, Schmidt, Schnabel, Schofield, Sch\"onbeck,
  Schreiber, Schuette, Schulte, Schutz, Schwalbe, Scott, Scott, Seidel,
  Sellers, Sengupta, Sentenac, Sequino, Sergeev, Setyawati, Shaddock, Shaffer,
  Shah, Shahriar, Shaner, Shao, Shapiro, Shawhan, Shen, Shoemaker, Shoemaker,
  Siellez, Siemens, Sieniawska, Sigg, Silva, Singer, Singh, Singhal, Sintes,
  Slagmolen, Slaven-Blair, Smith, Smith, Smith, Somala, Son, Sorazu,
  Sorrentino, Souradeep, Spencer, Srivastava, Staats, Steinke, Steinlechner,
  Steinlechner, Steinmeyer, Steltner, Stevenson, Stocks, Stone, Stops, Strain,
  Stratta, Strigin, Strunk, Sturani, Stuver, Summerscales, Sun, Sunil, Suresh,
  Sutton, Swinkels, Szczepa\ifmmode~\acute{n}\else \'{n}\fi{}czyk, Tacca, Tait,
  Talbot, Talukder, Tanner, T\'apai, Taracchini, Tasson, Taylor, Taylor,
  Tewari, Theeg, Thies, Thomas, Thomas, Thomas, Thorne, Thrane, Tiwari, Tiwari,
  Tokmakov, Toland, Tonelli, Tornasi, Torres-Forn\'e, Torrie, T\"oyr\"a,
  Travasso, Traylor, Trinastic, Tringali, Trozzo, Tsang, Tse, Tso, Tsuna,
  Tsukada, Tuyenbayev, Ueno, Ugolini, Urban, Usman, Vahlbruch, Vajente, Valdes,
  van Bakel, van Beuzekom, van~den Brand, Van Den~Broeck, Vander-Hyde, van~der
  Schaaf, van Heijningen, van Veggel, Vardaro, Varma, Vass, Vas\'uth, Vecchio,
  Vedovato, Veitch, Veitch, Venkateswara, Venugopalan, Verkindt, Vetrano,
  Vicer\'e, Viets, Vinciguerra, Vine, Vinet, Vitale, Vo, Vocca, Vorvick,
  Vyatchanin, Wade, Wade, Wade, Walet, Walker, Wallace, Walsh, Wang, Wang,
  Wang, Wang, Wang, Ward, Warner, Was, Watchi, Weaver, Wei, Weinert, Weinstein,
  Weiss, Wellmann, Wen, Wessel, We\ss{}els, Westerweck, Wette, Whelan, Whiting,
  Whittle, Wilken, Williams, Williams, Williamson, Willis, Willke, Wimmer,
  Winkler, Wipf, Wittel, Woan, Woehler, Wofford, Wong, Worden, Wright, Wu,
  Wysocki, Xiao, Yam, Yamamoto, Yancey, Yang, Yap, Yazback, Yu, Yu, Yvert,
  Zadro\ifmmode~\dot{z}\else \.{z}\fi{}ny, Zanolin, Zelenova, Zendri, Zevin,
  Zhang, Zhang, Zhang, Zhang, Zhang, Zhao, Zhou, Zhou, Zhu, Zhu, Zimmerman,
  Zlochower, Zucker, \& Zweizig}]{LIGOScientific:2018hze}
---. 2019, Phys. Rev. X, 9, 011001, \dodoi{10.1103/PhysRevX.9.011001}

\bibitem[{Al-Mamun {et~al.}(2021)Al-Mamun, Steiner, N\"attil\"a, Lange,
  O'Shaughnessy, Tews, Gandolfi, Heinke, \& Han}]{Al-Mamun:2020vzu}
Al-Mamun, M., Steiner, A.~W., N\"attil\"a, J., {et~al.} 2021, Phys. Rev. Lett.,
  126, 061101, \dodoi{10.1103/PhysRevLett.126.061101}

\bibitem[{Alford {et~al.}(2013)Alford, Han, \& Prakash}]{Alford:2013aca}
Alford, M.~G., Han, S., \& Prakash, M. 2013, Phys. Rev. D, 88, 083013,
  \dodoi{10.1103/PhysRevD.88.083013}

\bibitem[{Altiparmak {et~al.}(2022)Altiparmak, Ecker, \&
  Rezzolla}]{Altiparmak:2022bke}
Altiparmak, S., Ecker, C., \& Rezzolla, L. 2022, Astrophys. J. Lett., 939, L34,
  \dodoi{10.3847/2041-8213/ac9b2a}

\bibitem[{Annala {et~al.}(2022)Annala, Gorda, Katerini, Kurkela, N\"attil\"a,
  Paschalidis, \& Vuorinen}]{Annala:2021gom}
Annala, E., Gorda, T., Katerini, E., {et~al.} 2022, Phys. Rev. X, 12, 011058,
  \dodoi{10.1103/PhysRevX.12.011058}

\bibitem[{Annala {et~al.}(2020)Annala, Gorda, Kurkela, N\"attil\"a, \&
  Vuorinen}]{Annala:2019puf}
Annala, E., Gorda, T., Kurkela, A., N\"attil\"a, J., \& Vuorinen, A. 2020,
  Nature Phys., 16, 907, \dodoi{10.1038/s41567-020-0914-9}

\bibitem[{Annala {et~al.}(2018)Annala, Gorda, Kurkela, \&
  Vuorinen}]{Annala:2017llu}
Annala, E., Gorda, T., Kurkela, A., \& Vuorinen, A. 2018, Phys. Rev. Lett.,
  120, 172703, \dodoi{10.1103/PhysRevLett.120.172703}

\bibitem[{Antoniadis {et~al.}(2013)Antoniadis, Freire, Wex, Tauris, Lynch, van
  Kerkwijk, Kramer, Bassa, Dhillon, Driebe, Hessels, Kaspi, Kondratiev, Langer,
  Marsh, McLaughlin, Pennucci, Ransom, Stairs, van Leeuwen, Verbiest, \&
  Whelan}]{Antoniadis:2013pzd}
Antoniadis, J., Freire, P. C.~C., Wex, N., {et~al.} 2013, Science, 340,
  1233232, \dodoi{10.1126/science.1233232}

\bibitem[{Aoki {et~al.}(2006)Aoki, Endrodi, Fodor, Katz, \&
  Szabo}]{Aoki:2006we}
Aoki, Y., Endrodi, G., Fodor, Z., Katz, S.~D., \& Szabo, K.~K. 2006, Nature,
  443, 675, \dodoi{10.1038/nature05120}

\bibitem[{Bauswein {et~al.}(2019)Bauswein, Bastian, Blaschke, Chatziioannou,
  Clark, Fischer, \& Oertel}]{Bauswein:2018bma}
Bauswein, A., Bastian, N.-U.~F., Blaschke, D.~B., {et~al.} 2019, Phys. Rev.
  Lett., 122, 061102, \dodoi{10.1103/PhysRevLett.122.061102}

\bibitem[{Baym {et~al.}(1971)Baym, Pethick, \& Sutherland}]{Baym:1971pw}
Baym, G., Pethick, C., \& Sutherland, P. 1971, Astrophys. J., 170, 299,
  \dodoi{10.1086/151216}

\bibitem[{Capano {et~al.}(2020)Capano, Tews, Brown, Margalit, De, Kumar, Brown,
  Krishnan, \& Reddy}]{Capano:2019eae}
Capano, C.~D., Tews, I., Brown, S.~M., {et~al.} 2020, Nat. Astron., 4, 625,
  \dodoi{10.1038/s41550-020-1014-6}

\bibitem[{Cheng {et~al.}(2006)Cheng, Christ, Datta, van~der Heide, Jung,
  Karsch, Kaczmarek, Laermann, Mawhinney, Miao, Petreczky, Petrov, Schmidt, \&
  Umeda}]{Cheng:2006qk}
Cheng, M., Christ, N.~H., Datta, S., {et~al.} 2006, Phys. Rev. D, 74, 054507,
  \dodoi{10.1103/PhysRevD.74.054507}

\bibitem[{Cromartie {et~al.}(2019)Cromartie, Fonseca, Ransom,
  {et~al.}}]{NANOGrav:2019jur}
Cromartie, H.~T., Fonseca, E., Ransom, S.~M., {et~al.} 2019, Nature Astron., 4,
  72, \dodoi{10.1038/s41550-019-0880-2}

\bibitem[{Demorest {et~al.}(2010)Demorest, Pennucci, Ransom, Roberts, \&
  Hessels}]{Demorest:2010bx}
Demorest, P., Pennucci, T., Ransom, S., Roberts, M., \& Hessels, J. 2010,
  Nature, 467, 1081, \dodoi{10.1038/nature09466}

\bibitem[{Dietrich {et~al.}(2020)Dietrich, Coughlin, Pang, Bulla, Heinzel,
  Issa, Tews, \& Antier}]{Dietrich:2020efo}
Dietrich, T., Coughlin, M.~W., Pang, P. T.~H., {et~al.} 2020, Science, 370,
  1450, \dodoi{10.1126/science.abb4317}

\bibitem[{Drischler {et~al.}(2020)Drischler, Furnstahl, Melendez, \&
  Phillips}]{Drischler:2020hwi}
Drischler, C., Furnstahl, R.~J., Melendez, J.~A., \& Phillips, D.~R. 2020,
  Phys. Rev. Lett., 125, 202702, \dodoi{10.1103/PhysRevLett.125.202702}

\bibitem[{Drischler {et~al.}(2019)Drischler, Hebeler, \&
  Schwenk}]{Drischler:2017wtt}
Drischler, C., Hebeler, K., \& Schwenk, A. 2019, Phys. Rev. Lett., 122, 042501,
  \dodoi{10.1103/PhysRevLett.122.042501}

\bibitem[{Essick {et~al.}(2020)Essick, Landry, \& Holz}]{Essick:2019ldf}
Essick, R., Landry, P., \& Holz, D.~E. 2020, Phys. Rev. D, 101, 063007,
  \dodoi{10.1103/PhysRevD.101.063007}

\bibitem[{Essick {et~al.}(2021)Essick, Tews, Landry, \&
  Schwenk}]{Essick:2021kjb}
Essick, R., Tews, I., Landry, P., \& Schwenk, A. 2021, Phys. Rev. Lett., 127,
  192701, \dodoi{10.1103/PhysRevLett.127.192701}

\bibitem[{Fonseca {et~al.}(2021)Fonseca, Cromartie, Pennucci, Ray, Kirichenko,
  Ransom, Demorest, Stairs, Arzoumanian, Guillemot, Parthasarathy, Kerr,
  Cognard, Baker, Blumer, Brook, DeCesar, Dolch, Dong, Ferrara, Fiore,
  Garver-Daniels, Good, Jennings, Jones, Kaspi, Lam, Lorimer, Luo, McEwen,
  McKee, McLaughlin, McMann, Meyers, Naidu, Ng, Nice, Pol, Radovan,
  Shapiro-Albert, Tan, Tendulkar, Swiggum, Wahl, \& Zhu}]{Fonseca:2021wxt}
Fonseca, E., Cromartie, H.~T., Pennucci, T.~T., {et~al.} 2021, Astrophys. J.
  Lett., 915, L12, \dodoi{10.3847/2041-8213/ac03b8}

\bibitem[{Fujimoto {et~al.}(2023)Fujimoto, Fukushima, Hotokezaka, \&
  Kyutoku}]{Fujimoto:2022xhv}
Fujimoto, Y., Fukushima, K., Hotokezaka, K., \& Kyutoku, K. 2023, Phys. Rev.
  Lett., 130, 091404, \dodoi{10.1103/PhysRevLett.130.091404}

\bibitem[{Gal {et~al.}(2016)Gal, Hungerford, \& Millener}]{Gal:2016boi}
Gal, A., Hungerford, E.~V., \& Millener, D.~J. 2016, Rev. Mod. Phys., 88,
  035004, \dodoi{10.1103/RevModPhys.88.035004}

\bibitem[{Gerlach(1968{\natexlab{a}})}]{Gerlach:1968zz}
Gerlach, U.~H. 1968{\natexlab{a}}, Phys. Rev., 172, 1325,
  \dodoi{10.1103/PhysRev.172.1325}

\bibitem[{Gerlach(1968{\natexlab{b}})}]{Gerlach:thesis}
---. 1968{\natexlab{b}}, {Ph.D.~thesis, A third family of stable equilibria},
  Princeton University, (unpublished)

\bibitem[{Gorda {et~al.}(2023)Gorda, Komoltsev, \& Kurkela}]{Gorda:2022jvk}
Gorda, T., Komoltsev, O., \& Kurkela, A. 2023, Astrophys. J., 950, 107,
  \dodoi{10.3847/1538-4357/acce3a}

\bibitem[{Gorda {et~al.}(2021)Gorda, Kurkela, Paatelainen, S\"appi, \&
  Vuorinen}]{Gorda:2021kme}
Gorda, T., Kurkela, A., Paatelainen, R., S\"appi, S., \& Vuorinen, A. 2021,
  Phys. Rev. D, 104, 074015, \dodoi{10.1103/PhysRevD.104.074015}

\bibitem[{Han \& Steiner(2019)}]{Han:2018mtj}
Han, S., \& Steiner, A.~W. 2019, Phys. Rev. D, 99, 083014,
  \dodoi{10.1103/PhysRevD.99.083014}

\bibitem[{Hebeler {et~al.}(2013)Hebeler, Lattimer, Pethick, \&
  Schwenk}]{Hebeler:2013nza}
Hebeler, K., Lattimer, J.~M., Pethick, C.~J., \& Schwenk, A. 2013, Astrophys.
  J., 773, 11, \dodoi{10.1088/0004-637X/773/1/11}

\bibitem[{Hinderer(2008)}]{Hinderer:2007mb}
Hinderer, T. 2008, Astrophys. J., 677, 1216, \dodoi{10.1086/533487}

\bibitem[{Huth {et~al.}(2022)Huth, Pang, Tews, {et~al.}}]{Huth:2021bsp}
Huth, S., Pang, P. T.~H., Tews, I., {et~al.} 2022, Nature, 606, 276,
  \dodoi{10.1038/s41586-022-04750-w}

\bibitem[{Keller {et~al.}(2023)Keller, Hebeler, \& Schwenk}]{Keller:2022crb}
Keller, J., Hebeler, K., \& Schwenk, A. 2023, Phys. Rev. Lett., 130, 072701,
  \dodoi{10.1103/PhysRevLett.130.072701}

\bibitem[{Komoltsev \& Kurkela(2022)}]{Komoltsev:2021jzg}
Komoltsev, O., \& Kurkela, A. 2022, Phys. Rev. Lett., 128, 202701,
  \dodoi{10.1103/PhysRevLett.128.202701}

\bibitem[{Kurkela {et~al.}(2014)Kurkela, Fraga, Schaffner-Bielich, \&
  Vuorinen}]{Kurkela:2014vha}
Kurkela, A., Fraga, E.~S., Schaffner-Bielich, J., \& Vuorinen, A. 2014,
  Astrophys. J., 789, 127, \dodoi{10.1088/0004-637X/789/2/127}

\bibitem[{Kurkela {et~al.}(2010)Kurkela, Romatschke, \&
  Vuorinen}]{Kurkela:2009gj}
Kurkela, A., Romatschke, P., \& Vuorinen, A. 2010, Phys. Rev. D, 81, 105021,
  \dodoi{10.1103/PhysRevD.81.105021}

\bibitem[{Landry \& Essick(2019)}]{Landry:2018prl}
Landry, P., \& Essick, R. 2019, Phys. Rev. D, 99, 084049,
  \dodoi{10.1103/PhysRevD.99.084049}

\bibitem[{Landry {et~al.}(2020)Landry, Essick, \&
  Chatziioannou}]{Landry:2020vaw}
Landry, P., Essick, R., \& Chatziioannou, K. 2020, Phys. Rev. D, 101, 123007,
  \dodoi{10.1103/PhysRevD.101.123007}

\bibitem[{Leonhardt {et~al.}(2020)Leonhardt, Pospiech, Schallmo, Braun,
  Drischler, Hebeler, \& Schwenk}]{Leonhardt:2019fua}
Leonhardt, M., Pospiech, M., Schallmo, B., {et~al.} 2020, Phys. Rev. Lett.,
  125, 142502, \dodoi{10.1103/PhysRevLett.125.142502}

\bibitem[{Lim \& Holt(2022)}]{Lim:2022fap}
Lim, Y., \& Holt, J.~W. 2022.
\newblock \doarXiv{2204.09000}

\bibitem[{Linares {et~al.}(2018)Linares, Shahbaz, \& Casares}]{Linares:2018ppq}
Linares, M., Shahbaz, T., \& Casares, J. 2018, Astrophys. J., 859, 54,
  \dodoi{10.3847/1538-4357/aabde6}

\bibitem[{Lindblom(1998)}]{Lindblom:1998dp}
Lindblom, L. 1998, Phys. Rev. D, 58, 024008, \dodoi{10.1103/PhysRevD.58.024008}

\bibitem[{Lynn {et~al.}(2016)Lynn, Tews, Carlson, Gandolfi, Gezerlis, Schmidt,
  \& Schwenk}]{Lynn:2015jua}
Lynn, J.~E., Tews, I., Carlson, J., {et~al.} 2016, Phys. Rev. Lett., 116,
  062501, \dodoi{10.1103/PhysRevLett.116.062501}

\bibitem[{McLerran \& Reddy(2019)}]{McLerran:2018hbz}
McLerran, L., \& Reddy, S. 2019, Phys. Rev. Lett., 122, 122701,
  \dodoi{10.1103/PhysRevLett.122.122701}

\bibitem[{Miao {et~al.}(2020)Miao, Li, Zhu, \& Han}]{Miao:2020yjk}
Miao, Z., Li, A., Zhu, Z., \& Han, S. 2020, Astrophys. J., 904, 103,
  \dodoi{10.3847/1538-4357/abbd41}

\bibitem[{Miller {et~al.}(2020)Miller, Chirenti, \& Lamb}]{Miller:2019nzo}
Miller, M.~C., Chirenti, C., \& Lamb, F.~K. 2020, Astrophys. J., 888, 12,
  \dodoi{10.3847/1538-4357/ab4ef9}

\bibitem[{Miller {et~al.}(2019)Miller, Lamb, Dittmann, Bogdanov, Arzoumanian,
  Gendreau, Guillot, Harding, Ho, Lattimer, Ludlam, Mahmoodifar, Morsink, Ray,
  Strohmayer, Wood, Enoto, Foster, Okajima, Prigozhin, \&
  Soong}]{Miller:2019cac}
Miller, M.~C., Lamb, F.~K., Dittmann, A.~J., {et~al.} 2019, Astrophys. J.
  Lett., 887, L24, \dodoi{10.3847/2041-8213/ab50c5}

\bibitem[{Miller {et~al.}(2021)Miller, Lamb, Dittmann, Bogdanov, Arzoumanian,
  Gendreau, Guillot, Ho, Lattimer, Loewenstein, Morsink, Ray, Wolff, Baker,
  Cazeau, Manthripragada, Markwardt, Okajima, Pollard, Cognard, Cromartie,
  Fonseca, Guillemot, Kerr, Parthasarathy, Pennucci, Ransom, \&
  Stairs}]{Miller:2021qha}
---. 2021, Astrophys. J. Lett., 918, L28, \dodoi{10.3847/2041-8213/ac089b}

\bibitem[{Most {et~al.}(2019)Most, Papenfort, Dexheimer, Hanauske, Schramm,
  St\"ocker, \& Rezzolla}]{Most:2018eaw}
Most, E.~R., Papenfort, L.~J., Dexheimer, V., {et~al.} 2019, Phys. Rev. Lett.,
  122, 061101, \dodoi{10.1103/PhysRevLett.122.061101}

\bibitem[{Most {et~al.}(2018)Most, Weih, Rezzolla, \&
  Schaffner-Bielich}]{Most:2018hfd}
Most, E.~R., Weih, L.~R., Rezzolla, L., \& Schaffner-Bielich, J. 2018, Phys.
  Rev. Lett., 120, 261103, \dodoi{10.1103/PhysRevLett.120.261103}

\bibitem[{Raaijmakers {et~al.}(2020)Raaijmakers, Greif, Riley,
  {et~al.}}]{Raaijmakers:2019dks}
Raaijmakers, G., Greif, S.~K., Riley, T.~E., {et~al.} 2020, Astrophys. J.
  Lett., 893, L21, \dodoi{10.3847/2041-8213/ab822f}

\bibitem[{Raaijmakers {et~al.}(2021)Raaijmakers, Greif, Hebeler, Hinderer,
  Nissanke, Schwenk, Riley, Watts, Lattimer, \& Ho}]{Raaijmakers:2021uju}
Raaijmakers, G., Greif, S.~K., Hebeler, K., {et~al.} 2021, Astrophys. J. Lett.,
  918, L29, \dodoi{10.3847/2041-8213/ac089a}

\bibitem[{Riley {et~al.}(2019)Riley, Watts, Bogdanov, Ray, Ludlam, Guillot,
  Arzoumanian, Baker, Bilous, Chakrabarty, Gendreau, Harding, Ho, Lattimer,
  Morsink, \& Strohmayer}]{Riley:2019yda}
Riley, T.~E., Watts, A.~L., Bogdanov, S., {et~al.} 2019, Astrophys. J. Lett.,
  887, L21, \dodoi{10.3847/2041-8213/ab481c}

\bibitem[{Riley {et~al.}(2021)Riley, Watts, Ray, Bogdanov, Guillot, Morsink,
  Bilous, Arzoumanian, Choudhury, Deneva, Gendreau, Harding, Ho, Lattimer,
  Loewenstein, Ludlam, Markwardt, Okajima, Prescod-Weinstein, Remillard, Wolff,
  Fonseca, Cromartie, Kerr, Pennucci, Parthasarathy, Ransom, Stairs, Guillemot,
  \& Cognard}]{Riley:2021pdl}
Riley, T.~E., Watts, A.~L., Ray, P.~S., {et~al.} 2021, Astrophys. J. Lett.,
  918, L27, \dodoi{10.3847/2041-8213/ac0a81}

\bibitem[{Romani {et~al.}(2022)Romani, Kandel, Filippenko, Brink, \&
  Zheng}]{Romani:2022jhd}
Romani, R.~W., Kandel, D., Filippenko, A.~V., Brink, T.~G., \& Zheng, W. 2022,
  Astrophys. J. Lett., 934, L18, \dodoi{10.3847/2041-8213/ac8007}

\bibitem[{{Schaeffer} {et~al.}(1983){Schaeffer}, {Zdunik}, \&
  {Haensel}}]{Schaeffer_PT_destab}
{Schaeffer}, R., {Zdunik}, L., \& {Haensel}, P. 1983, \aap, 126, 121.
\newblock \url{https://ui.adsabs.harvard.edu/abs/1983A&A...126..121S}

\bibitem[{Schertler {et~al.}(2000)Schertler, Greiner, Schaffner-Bielich, \&
  Thoma}]{Schertler:2000xq}
Schertler, K., Greiner, C., Schaffner-Bielich, J., \& Thoma, M.~H. 2000, Nucl.
  Phys. A, 677, 463, \dodoi{10.1016/S0375-9474(00)00305-5}

\bibitem[{{Seidov}(1971)}]{Seidov_PT_destab}
{Seidov}, Z.~F. 1971, \sovast, 15, 347.
\newblock \url{https://ui.adsabs.harvard.edu/abs/1971SvA....15..347S}

\bibitem[{Somasundaram {et~al.}(2023)Somasundaram, Tews, \&
  Margueron}]{Somasundaram:2021clp}
Somasundaram, R., Tews, I., \& Margueron, J. 2023, Phys. Rev. C, 107, 025801,
  \dodoi{10.1103/PhysRevC.107.025801}

\bibitem[{Tews {et~al.}(2013)Tews, Kr\"uger, Hebeler, \& Schwenk}]{Tews:2012fj}
Tews, I., Kr\"uger, T., Hebeler, K., \& Schwenk, A. 2013, Phys. Rev. Lett.,
  110, 032504, \dodoi{10.1103/PhysRevLett.110.032504}

\bibitem[{Tews {et~al.}(2018)Tews, Margueron, \& Reddy}]{Tews:2018iwm}
Tews, I., Margueron, J., \& Reddy, S. 2018, Phys. Rev. C, 98, 045804,
  \dodoi{10.1103/PhysRevC.98.045804}

\end{thebibliography}

\end{document}